\documentclass[twocolumn,aps,pra,superscriptaddress,showpacs]{revtex4}
\usepackage{graphicx,dcolumn}
\usepackage{amsfonts,amssymb,amsmath}
\usepackage{ulem}
\usepackage[english]{babel}
\newcommand{\leqsi}{\; {\scriptstyle {< \atop \sim}} \;}

\begin{document}

\title{Coherent propagation of waves in random media with weak nonlinearity}
\author{Thomas Wellens}
\affiliation{Physikalisches Institut, Albert-Ludwigs Universit\"at
  Freiburg, Hermann-Herder-Str. 3, 79104 Freiburg, Germany}
\author{Beno\^it Gr\'emaud}
\affiliation{Centre for Quantum Technologies, 
National University of Singapore, 3 Science Drive 2, Singapore 117543, 
Singapore}
\affiliation{Laboratoire Kastler-Brossel, UPMC-Paris 6, ENS, CNRS;
4 Place Jussieu, F-75005 Paris, France}

\date{\today}
\begin{abstract}
We develop a diagrammatic theory for transport of waves in disordered media with weak nonlinearity.
We first represent the solution of the nonlinear wave equation as a nonlinear Born series. From this, we construct nonlinear ladder and crossed diagrams for the average wave intensity. Then, we sum up the  diagrammatic series completely, i.e. nonperturbatively in the strength of the nonlinearity, and thereby obtain integral equations describing both nonlinear diffusive transport and coherent backscattering of the average intensity. As main result, we find that the nonlinearity significantly influences the magnitude of the coherent backscattering effect. Depending on the
type of nonlinearity, coherent backscattering is either enhanced or suppressed, as compared to the linear case.
\end{abstract}

\maketitle

\section{Introduction}

Multiple scattering of waves in random media has been an active area of
research for about hundred years. Of special interest is the role of interferences between different multiple scattering paths. For a long time, it was believed that
all interferences are destroyed by the disorder, such that the propagation of
the average intensity follows a simple diffusion process. But in 1958, Anderson
showed that interference effects can lead to a localization of wavefunctions
in certain random lattices, corresponding to a complete suppression of diffusion \cite{anderson58}. 
Later, it was recognized that even in weakly scattering media,
interferences between amplitudes propagating along reversed scattering paths
may survive the ensemble average \cite{langer66}, leading to the
phenomena of weak
localization (reduction of the diffusion constant) and coherent
backscattering (enhancement of the average intensity scattered into the direction
exactly opposite to the incident wave) \cite{kuga,albada,wolf}.

Whereas these mesoscopic physics effects  have been thoroughly studied for the case of linear wave equations,
important, yet unresolved questions concern the impact of interactions or nonlinearities on
the multiple scattering interferences. For example, does  Anderson localization still persist in the
presence of nonlinearities? 
Good candidates for studying these questions appear to be
Bose-Einstein condensates in disordered potentials \cite{bec}. In the mean field regime, the condensate is still described by a single coherent wave function following a nonlinear wave equation (Gross-Pitaevskii equation). Thus, the coherence of 
the condensate is preserved in presence of (not too strong) interactions. Similar nonlinear equations also describe propagation of light in disordered nonlinear media \cite{skipetrov,spivak}. In contrast, the situation is quite different in the
case of electronic transport \cite{altshuler99}, where the interactions combined with finite temperature effects give rise to dephasing mechanisms, which in general destroy the disorder-induced coherent effects.

As an important step in the theoretical description of coherent effects in nonlinear disordered systems,
a theory for coherent backscattering in presence of nonlinearity has recently been developed \cite{wellens08,wellens09}. 
Most importantly, it has been shown that the usual picture of coherent backscattering
resulting from interference between pairs of only two counterpropagating
amplitudes breaks down in the
nonlinear regime. As a consequence of 
nonlinear coupling between
different partial waves, nonlinear coherent backscattering must rather be interpreted
as a multi-wave interference phenomenon \cite{wellens05,wellens06b}.  This fundamental property was not taken into account
in earlier works \cite{agranovich91,heiderich95}, which therefore
concluded that nonlinearity would leave the enhancement factor
in exact backscattering direction unchanged with respect to the linear case - in contradiction to our results presented below.

 The present paper provides a complete and detailed presentation of the diagrammatic
 theory sketched in \cite{wellens08}, which fully takes into account the multi-wave interference character of nonlinear coherent backscattering.
 To perform the disorder average, we will define nonlinear ladder and crossed diagrams, which describe diffusive transport and coherent backscattering of the average wave intensity in the dilute regime $k\ell\gg 1$ (with wavenumber $k$ and mean free path $\ell$).
 We are able to sum up the diagrammatic series completely, i.e. nonperturbatively in the strength of the nonlinearity.
As final result, we will see that even very weak nonlinearities do significantly affect the
coherent backscattering enhancement factor, which, depending on the type of nonlinearity, may be either enhanced or reduced, as compared to the linear case. These predictions of our theory agree with recent numerical simulations \cite{wellens08,hartung08}.

 The outline of the paper is as follows. In Sec.~\ref{setup}, we describe the general conditions under which the nonlinear diagrammatic theory can be applied. In short, these are the following: first, we consider 
a nonlinear equation for the complex wave amplitude $E({\bf r})$, which we assume to possess a unique stationary and stable solution. 
In order to perform the disorder average,
we require, second, the disordered medium to be dilute. Finally, we neglect nonlinear effects which become relevant only at larger values of the nonlinearity. 

In the present paper, we will develop our theory on the basis of two different physical models, corresponding to homogeneous or discrete
nonlinearity, respectively.
In Sec.~\ref{moddiag}, we start by considering the discrete case, i.e.
a collection of nonlinear point scatterers placed at random positions. We obtain the formal solution of the wave equation
as a nonlinear Born series, and introduce diagrams representing the individual terms of this series.
Furthermore, we identify certain types of diagrams  (ladder and crossed diagrams) which survive the disorder average over the random positions of scatterers.

Among these, the ladder diagrams describe incoherent nonlinear transport of the average wave intensity. This is the subject of Sec.~\ref{sladder}. We analyze the statistical fluctuations of the wave intensity (which influence the average intensity, since the nonlinearity couples different statistical moments), define
refractive indices for scattered and incident waves, and finally arrive at a set of
 nonlinear integral equations for the average intensity inside the scattering volume. 

In Sec.~\ref{scrossed}, we analyze the crossed diagrams, which describe the coherent backscattering effect as interferential correction to incoherent transport.
We identify the building blocks for the nonlinear crossed diagrams, and point out the rules according to which these elements are connected to each other.
This leads us to integral equations for the crossed intensity, which finally determine the
coherent backscattering enhancement factor. We will discuss under which conditions the nonlinearity reduces or enhances the coherent backscattering effect, as compared to the linear case.

In Sec.~\ref{skerr},
we consider  a second physical model, namely a disordered collection of linear scatterers embedded into a homogeneous
nonlinear Kerr medium.
 Equivalently, the same equation also describes Bose-Einstein condensates in disordered potentials (Gross-Pitaevskii equation).
For this model, our theory can be applied in almost the same way as for the nonlinear scatterer model - in particular, the
finally obtained average transport equations turn out to have the same form. We solve these equations numerically for different values of the
Kerr nonlinearity constant $\alpha$, and thereby confirm the general considerations of Sec.~\ref{scrossed}. 

Finally, Sec.~\ref{concl} concludes the paper. Some more technical calculations are relegated to two appendices. App.~\ref{sctrans} contains the complete
set of crossed transport equations, as they result from the exact summation of all crossed diagrams. In App.~\ref{proofrel2}, we prove a general relation between diffuson and cooperon cross section, which is useful in order to determine whether the nonlinearity decreases or enhances the coherent backscattering effect.

\section{General setting}
\label{setup}

In this section, we will expose the general assumptions which we will need in 
Secs.~\ref{moddiag}-\ref{scrossed} for developing the nonlinear diagrammatic theory.

\subsection{Stationary state}
\label{ssstable}

First, we assume that our system is described by a complex wave function ${\tilde E}({\bf r},t)\in {\mathbb C}$, which fulfills a nonlinear random wave equation. Furthermore, we assume that, for (almost) any realization of the disorder, the field 
${\tilde E}({\bf r},t)$
approaches, for $t\to\infty$, a stationary state of the form ${\tilde E}({\bf r},t)=e^{i\omega t}E({\bf r})$,
where the frequency $\omega$ is determined by the external, monochromatic source. Whereas for a linear system, this assumption implies no
loss of generality, since the time-dependent behavior for arbitrary (non-monochromatic) sources follows simply by Fourier transformation with respect to
$\omega$, the same is not true in the nonlinear case. For example, it has been shown
that solutions of the nonlinear Gross-Pitaevskii equation exhibit permanently time-dependent behavior
even if the external source is purely monochromatic \cite{paul05}. For a system of finite size, however, we expect
the assumption of a monochromatic stationary solution to be valid if the nonlinearity is small enough,
and if generation of higher harmonics can be neglected. Using the language of nonlinear optics,
the latter condition implies that
all nonlinear susceptibilities of even order (i.e. $\chi^{(2)}$, $\chi^{(4)}$,\dots), which appear for noncentrosymmetric media \cite{boyd}, vanish.
Apart from that, the polarizability of the nonlinear medium may be an arbitrary function of the local intensity $|E({\bf r})|^2$.

Under these conditions, the field $E({\bf r})$ fulfills a stationary, nonlinear wave equation, see for example Eq.~(\ref{model1}) or Eq.~(\ref{hom}) below. This wave equation will serve as the starting point for our diagrammatic theory, see Sec.~\ref{moddiag}. Let us note however, that the diagrammatic theory in its present form is not able  to predict the maximum strength of the nonlinearity up to which the stationary assumption remains valid. This must be checked by alternative methods, for example
direct numerical simulations. For the case of the homogeneous Kerr nonlinearity, see Eq.~(\ref{hom}),
a theoretical prediction of the instability threshold was performed in \cite{skipetrov}. Numerical simulations for nonlinear point 
scatterers \cite{gremaud08} and the 2D Gross-Pitaevskii equation \cite{hartung08} indicate that, in these cases, the instability threshold is large enough 
such that important nonlinear effects (i.e. modification or even inversion of the coherent backscattering effect) are observed in the stable regime, for which our theory applies.

\subsection{Dilute medium}

The second assumption required for our diagrammatic theory is the validity of a dilute medium
approximation. When calculating the average scattered intensity, this assumption allows us to discard
all diagrams except so-called ladder and crossed diagrams. 
From the ladder diagrams, we then obtain the diffuse background
of the average scattered intensity, whereas the crossed diagrams yield the interference contribution leading to coherent backscattering. In the linear case, this dilute medium approximation is known to be valid if 
$k\ell\gg 1$, i.e. if the mean free path  $\ell$ is much larger than the wavelength $\lambda=2\pi/k$
divided by $2\pi$. We now assume the phase randomization mechanism, which makes it possible
 to neglect all diagrams except ladder and crossed diagrams, to be equally effective also in the nonlinear case. In Sec.~\ref{sdiag}, we will then define nonlinear generalizations of
 ladder and crossed diagrams, which we sum up in order to obtain
 nonlinear transport equations, see Secs.~\ref{sladder} and \ref{scrossed}.
 
 We  note that, in principle, the dilute medium condition $k\ell\gg 1$ might be modified
 by the presence of the nonlinearity. This must be checked for the specific model under consideration. 
 For example, it can be shown that the diagrams describing scattering processes induced by the nonlinear index fluctuations
 $\alpha|E({\bf r})|^2$ in the case of the homogeneous Kerr nonlinearity, see Eq.~(\ref{hom}), are neither of
 ladder nor crossed type. Since these processes depend on the strength $\alpha$ of the nonlinearity,
 the condition $k\ell\gg 1$ alone is not sufficient to guarantee their negligibility. Instead, an additional assumption for $\alpha$ is required,
 see Sec.~\ref{smodel2}. On the other hand, for the nonlinear point scatterer model, see Sec.~\ref{smodel1}, where the nonlinearity is concentrated inside the scatterers, we expect the diluteness condition to be fulfilled for arbitrary strength of the nonlinearity, if the average distance between the scatterers is large enough.

\subsection{Neglect of nonlinear self-action effects}
\label{setupsa}

Finally, our last assumption concerns the neglect of nonlinear self-action effects (self-focusing, filamentation,...) originating from the intensity-dependent refractive index \cite{boyd}.
We assume those effects to be negligible on the length scale of the mean free path $\ell$.
More precisely, we approximate the optical path between two scattering events
at ${\bf r_1}$ and ${\bf r_2}$, with $|{\bf r_1}-{\bf r_2}|\simeq \ell$, by a straight line, see Eq.~(\ref{green}),
which also means that the spherical geometry of field radiated by a point source is preserved.
This is justified if the intensity-dependent refractive index $n({\bf r})$ varies slowly on the
length scale $\ell$, i.e. $\ell |\nabla n({\bf r})|\ll 1$. In principle, these
effects could be taken into account within the diagrammatic approach. The determination of the exact optical
path required for this purpose, however, would enormously increase the computational effort.

In summary, all the three above general conditions for the applicability of our diagrammatic approach
are fulfilled in the case of dilute media and weak nonlinearity. Let us stress, however, that
our approach is nevertheless non-perturbative in the strength $\alpha$ of the nonlinearity. As we will see later,
even a very weak nonlinearity can have a large impact on the final result
(i.e. the height of the coherent backscattering cone), if the size of the system (optical thickness)
is large. In order to access the regime where
perturbative expansions in powers of $\alpha$  do not rapidly converge (even if $\alpha$ is small),
we will therefore have to sum up 
our diagrammatic series completely, without restriction to certain orders of $\alpha$.

\section{Model and diagrams}
\label{moddiag}

In Sec.~\ref{setup}, we have discussed general conditions for our diagrammatic approach.
In order to present the theory in a concrete form,
we now specify the nonlinear wave equation under consideration,
and define a concrete model for the disorder.

In this paper, 
we will consider
the following two physical models: a collection of nonlinear point scatterers, see Sec.~\ref{smodel1}, and a
collection of linear scatterers embedded in a homogeneous Kerr medium, see Sec.~\ref{smodel2}. 
In the second model, the Kerr nonlinearity induces an intensity-dependent refractive index, which
modifies the average wave propagation between two linear scattering events.
Similarly, also in the first model, the nonlinear point scatterers form an
effective nonlinear medium, but - in addition to that - also the
scattering events are nonlinear. In this sense, the nonlinear point scatterer model
is the more general one, and we will therefore use it as our starting point for the
diagrammatic derivation of nonlinear transport equations. The modifications necessary for
the homogeneous nonlinearity will be discussed in Sec.~\ref{smodel2}.

\subsection{Description of the model: nonlinear point scatterers}
\label{smodel1}

We consider an assembly of $N$ point-like
scatterers located at randomly chosen 
positions $\mathbf{r}_i$, $i=1,\dots,N$ inside a sample volume $V$
illuminated by a plane wave $\mathbf{k}_L$. We assume the field
radiated by each scatterer to be a nonlinear function $f(E_i)$
of the local field $E_i=E({\bf r_i})$.
Since we neglect higher harmonics, the radiated field must attain the form
\begin{equation}
f(E)=g(I)E
\end{equation}
where
$I=EE^*$ is the local intensity, and $g(I)$ is proportional to the
polarizability of the scatterers, which may be an arbitrary (smooth) function of $I$.
This results in a set of
nonlinear equations for the field at each scatterer:
\begin{equation}
\label{model1}
E_i=e^{i\mathbf{k}_L\cdot \mathbf{r}_i}+\sum_{j\neq i} G_{ij} f(E_j)
\end{equation}
where 
\begin{equation}
 G_{ij}=G_0({\bf r_i},{\bf r_j})=\frac{e^{ik|{\bf r}_i-{\bf
  r}_j|}}{4\pi|{\bf r}_i-{\bf r}_j|}\label{green0}
\end{equation} 
with $k=|{\bf  k}_L|$, denotes the vacuum Green's
function. Note that the incident plane wave is normalized to unit intensity, which can always be achieved by a suitable redefinition of $g(I)$.

In the case of conservative scatterers, the optical theorem 
imposes the following condition:
\begin{equation}
{\rm Im}g(I)=\frac{k}{4\pi}|g(I)|^2\label{optth}
\end{equation}
which has a general solution of the form
\begin{equation}
g(I)=\frac{4\pi i}{k}\frac{e^{2i\delta(I)}+1}{2}\label{gcons}
\end{equation}
with phase shift $\delta(I)\in {\mathbb R}$. Let us note, however, that our diagrammatic approach is valid for arbitrary polarizability $g(I)$. Hence, we will also consider amplifying or absorbing scatterers in the following,
where Eqs.~(\ref{optth},\ref{gcons}) are not fulfilled.

Finally, the field $E_D$ measured by a far-field detector placed
in direction $\mathbf{k_D}$ (with $|\mathbf{k_D}|=k$) is given by
\begin{equation}
E_D({\bf k_D})= \sum_i \frac{e^{-i\mathbf{k_D}\cdot \mathbf{r}_i}}{4\pi R} f(E_i)\label{ed}
\end{equation}
where $R$ denotes the distance to the detector.
From $E_D$, the intensity scattered into direction $\mathbf{k_D}$, measured as a dimensionless quantity (independent of $R$), the so-called "bistatic coefficient" \cite{ishimaru}, results as 
\begin{equation}
\Gamma({\bf k_D})=\frac{4\pi R^2}{A} |E_D({\bf k_D})|^2\label{gamma}
\end{equation}
where $A$ denotes the transverse area (with respect to the incoming laser beam) of the sample volume $V$. 

\subsection{Diagrammatic representation}
\label{sdiag}

Our aim is to  calculate the mean scattered intensity $\langle \Gamma({\bf k_D})\rangle$, averaged over the random positions of scatterers. For this purpose, we employ a method
based on the diagrammatic representation of the iterative solution of the nonlinear
multiple scattering equations (\ref{model1}).

\subsubsection{A single nonlinear scattering event}
\label{sssingle}

\begin{figure}
\centerline{\includegraphics[width=8cm]{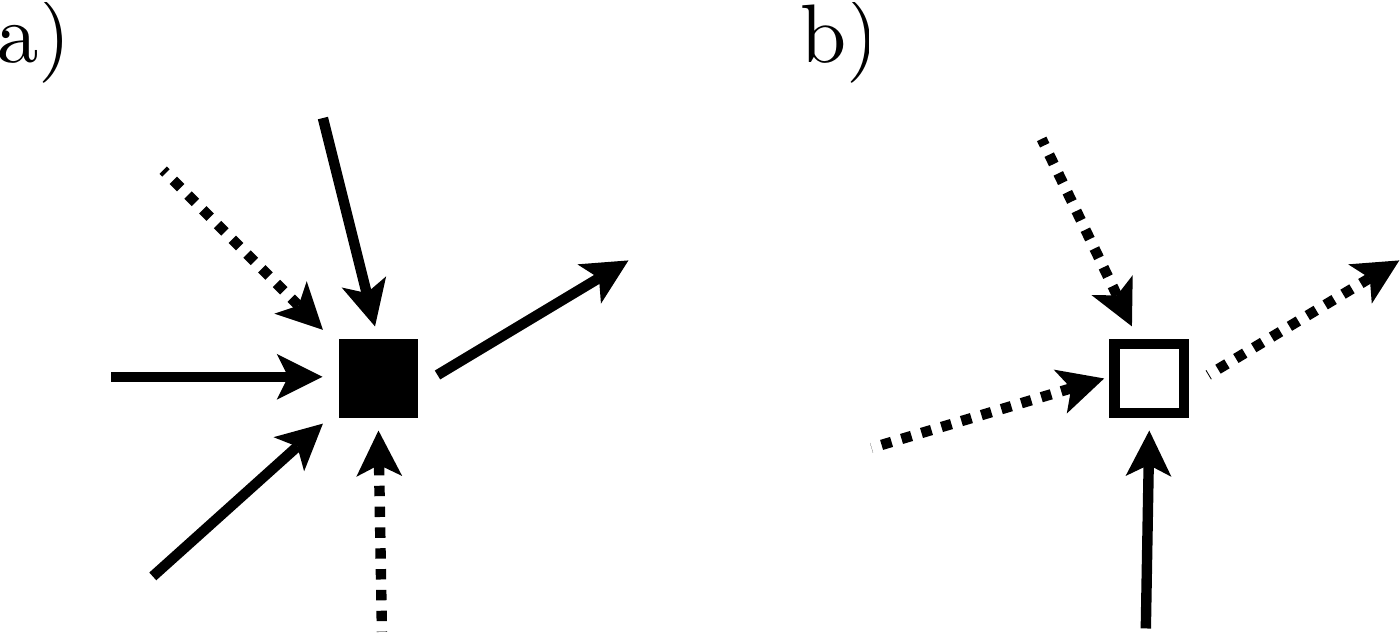}}
\caption{Diagrammatic representation of nonlinear scattering events.
The left half (a) represents 
a scattering event of nonlinear order $n=2$, i.e. 
the term $g_2 E^3(E^*)^2$ in the series (\ref{gn}) for the scattered field $f(E)$.
The right half (b) represents a complex conjugate event of order $n=1$, i.e. the term $g_1^* (E^*)^2 E$ 
in the complex conjugate series for $f^*(E)$.
\label{single}} 
\end{figure}

First, let us examine a single event, described by the function $f(E)=g(I)E$.
Expanding the polarizability $g(I)=\sum_n g_n I^n$ in a Fourier series, we obtain the following
sum over the order $n$ of the nonlinearity:
\begin{equation}
f(E)=\sum_{n=0}^\infty g_n E^{n+1}\left(E^*\right)^n\label{gn}
\end{equation}
The terms of this sum
are represented by diagrams like 
Fig.~\ref{single}, which shows, as an example, $f$ for $n=2$ and $f^*$ for $n=1$.
In Fig.~\ref{single}(a), the outgoing solid arrow represents the vacuum
  Green's function $G_{ij}$ pointing from position
  $\bf r_j$, where the scattering event $f$ occurs, to position $\bf
  r_i$, where the next scattering event will occur in the subsequent
  iteration step, see below. The incoming solid and dashed arrows represent the
  arguments $E_j$ and $E_j^*$ of the nonlinear function $f$, each of which may have undergone different
  previous scattering events. Due to
  the neglect of higher harmonics,
  the number of incoming solid arrows must exceed the number of dashed arrows by
  one. Of course, the opposite is true for the complex conjugate $f^*$, Fig.~\ref{single}(b). 
  
\subsubsection{Many scattering events: nonlinear Born series for the wave amplitude}
\label{snlfield}
  
\begin{figure}
\centerline{\includegraphics[width=8cm]{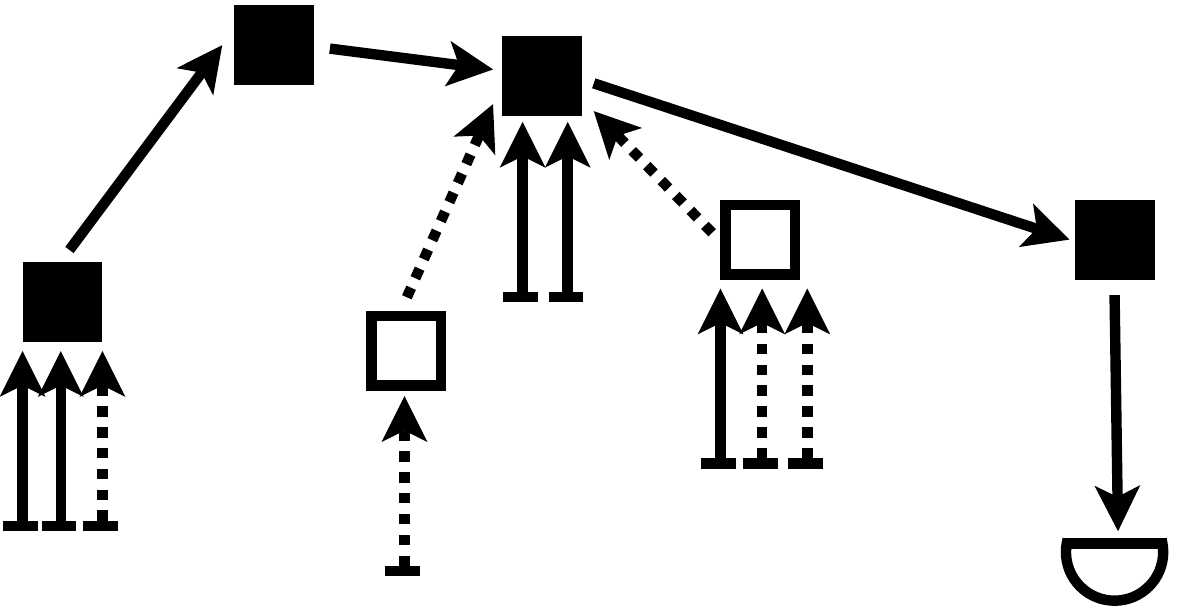}}
\caption{Example of a nonlinear field diagram, obtained by
  iteration of the
multiple scattering equation, Eq.~(\ref{model1}). The squares $\blacksquare$ (or $\Box$) represent scattering events  $f$ (or $f^*$). Solid (or dashed) arrows connecting two squares represent vacuum Green's functions $G_{ij}$ (or $G_{ij}^*$), and arrows with bar the  incident plane wave $\bf k_L$. Finally, the half-circle symbolizes the detector placed in direction $\bf k_D$, see Eq.~(\ref{ed}).
\label{multiple}} 
\end{figure}

Using single scattering events as building blocks, we can now compose diagrams for
the wave amplitude $E_D$.
Formally, a solution of Eq.~(\ref{model1}) is
obtained by iteration: we start with $E_i^{(0)}=0$, $i=1,\dots N$
on the right hand side, and obtain
$E_i^{(1)}=\exp(i\mathbf{k}_L\cdot \mathbf{r}_i)$ on the left hand
side. This result again inserted on the right hand side yields $E_i^{(2)}$, and so on.
In this way, the solution of Eq.~(\ref{model1}) is obtained as a nonlinear Born series, i.e. a sum over sequences of single nonlinear scattering events. The individual terms of this sum are represented by diagrams of
the form shown in Fig.~\ref{multiple}. 
Here, all the arrows connecting two squares correspond to vacuum Green's functions $G_{ij}$, whereas the remaining ones denote incident or outgoing waves.

Please notice the following general structure of all nonlinear field diagrams: Every full square 
($\blacksquare$) of nonlinear order $n$ exhibits exactly one outgoing solid arrow, $n$ incoming dashed, and $n+1$ incoming solid arrows. 
Likewise, every open square ($\Box$) has one outgoing dashed arrow, $n$ incoming solid, and $n+1$ incoming dashed arrows.  

Note that, for simplicity, we have omitted in Fig.~\ref{multiple} any indices labelling the different scatterers. The reader should simply imagine that every single event ($\blacksquare$ or $\Box$) occurs at some given scatterer $i$.  It may be that  two events occur at the same scatterer  - only neighboring squares (which are connected by a single arrow) must  correspond to different scatterers, since only couplings $G_{ij}$ between different scatterers $i\neq j$
occur in Eq.~(\ref{model1}).  (Repeated scattering by the same scatterer is already included in the polarizability function $g$).

We note that the diagrammatic representation of the nonlinear Born series has already been examined in a previous article \cite{wonderen}
(for the special case of a $\chi^{(3)}$-nonlinearity), which, however, does not make use of the dilute medium approximation, as we do in the following.

\subsubsection{Nonlinear ladder and crossed diagrams for the average wave intensity}
\label{ssladder}
\label{sscrossed}

Next, we consider diagrams for the detected intensity
$\Gamma\propto E_DE_D^*$, see Eq.~(\ref{gamma}), which
consist of one diagram representing $E_D$ and another one representing $E_D^*$, and 
perform the ensemble average over the random positions ${\bf r}_1,\dots,{\bf r}_N$ of the scatterers.

For this purpose, we will employ
the dilute medium approximation $k\ell\gg 1$, where $\ell$ denotes the mean
free path, see below.
From the linear case, it is well known that - in
leading order of $1/(k\ell)$ - only the so-called
ladder and crossed diagrams contribute to the average intensity
$\langle E_D E_D^*\rangle$. The reason why those diagrams survive the
ensemble average is that, in the exact backscattering direction, ${\bf k_D}=-{\bf k_L}$,
all the complex phase factors contained in the vacuum Green's
functions $G_{ij}$ cancel each other. (In
other directions ${\bf k_D}\neq -{\bf k_L}$, this is still true for the ladder,
but not for the crossed diagrams.)
This disappearance of phase factors is achieved by grouping together
solid and
dashed arrows, such
that for each vacuum Green's function $G_{ij}$ there exists a corresponding
$G_{ij}^*$ or $G_{ji}^*$ which cancels the phase factor. Since $G_{ij}=G_{ji}$ (reciprocity symmetry), the
solid and dashed arrows may point in the same or in opposite
direction, giving rise to ladder or crossed diagrams, respectively.

\begin{figure}
\centerline{\includegraphics[width=8cm]{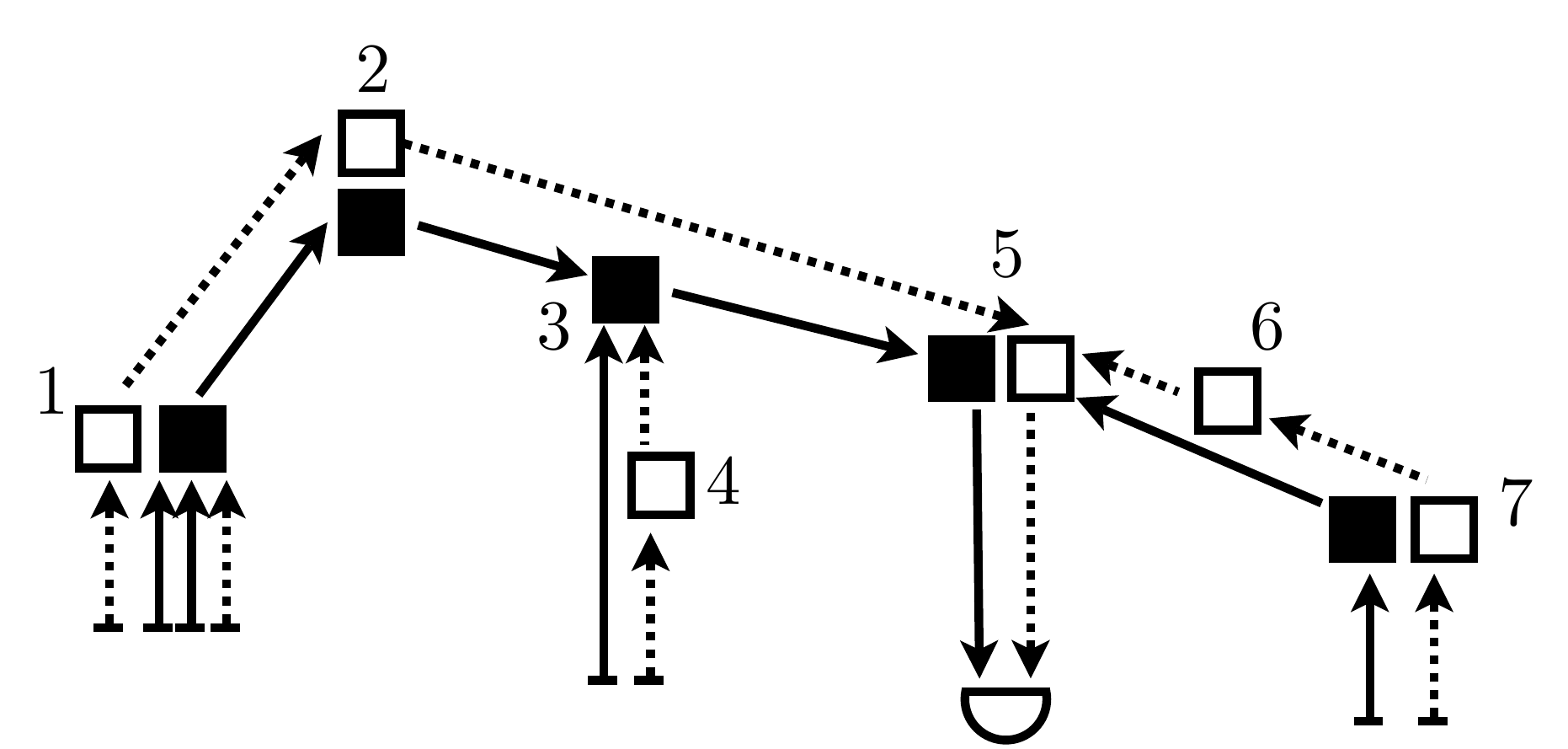}}
\caption{Example of a nonlinear ladder diagram. It is obtained from two nonlinear field diagrams for $E_D$ and $E_D^*$, respectively (see Fig.~\ref{multiple}), by grouping together the nonlinear events $\blacksquare$ and $\Box$ such that every vacuum Green's function $G_{ij}$ (solid arrow) is accompanied by a complex conjugate $G_{ij}^*$ (dashed arrow).  Scattering events (see Sec.~\ref{sscatt}) occur at scatterer 1, 2, 5, and 7, where  $\blacksquare$ and $\Box$ form a pair, whereas scatterer 3, 4, and 6 contribute to average propagation (see Sec.~\ref{sindex}).
\label{ladder}} 
\end{figure}

In the following, we will now apply the same idea to our nonlinear diagrams. A ladder diagram is obtained if all the vacuum Green's functions occurring in the diagram for $E_DE_D^*$ are grouped together in pairs $G_{ij} G^*_{ij}$  pointing in the same direction. An example is shown in
Fig.~\ref{ladder}.
 Here, we adopt the convention that two squares ($\blacksquare$ and $\Box$) drawn next to each other correspond to the same scatterer $i$ (thus forming a scattering event at $i$, see below), whereas all other events occur at different scatterers $j\neq i$. Hence, the diagram shown in Fig.~\ref{ladder} involves $7$ different scatterers. Moreover, since the ladder diagrams describe the average intensity, an integration over the positions ${\bf r_i}$ of the scatterers is implicitly assumed. Thus, the example shown in Fig.~\ref{ladder} represents the following expression:
\begin{eqnarray}
& & {\mathcal N}^7 g_0^3 \left(g_0^*\right)^5 g_1^2 g_1^*\int d{\bf r_1}\dots d{\bf r_7} e^{i{\bf k_L}\cdot ({\bf r_3}-{\bf r_4})}\nonumber\\
& &  \left|G_{12}\right|^2 G_{23}G_{25}^*G_{43}^*G_{35}G_{65}^*G_{76}^*G_{75}\label{ladderexample}
\end{eqnarray}
where ${\mathcal N}=N/V$ denotes the density of scatterers.
Note that the phase factors $\exp(\pm i{\bf k_L}\cdot {\bf r_{1,7}})$ representing the waves incident on scatterers 1 and 7, and the waves $\exp(\mp i{\bf k_D}\cdot {\bf r_5})$ emitted from scatterer 5 cancel each other, and are hence not included in Eq.~(\ref{ladderexample}).

The example shown in  Fig.~\ref{ladder}  is also useful to understand  the difference between scattering events for the average intensity, on the one hand, and events contributing to average field propagation, on the other hand. Scattering events always occur as a pair of two single events
($\blacksquare$ and $\Box$), i.e., at scatterer 
1,2,5, and 7 in Fig.~\ref{ladder}, where the wave amplitude $E$ and its complex conjugate $E^*$ are scattered
by the same scatterer, contributing thus to the propagation of the average intensity
$\langle EE^*\rangle$.
In contrast, the events at $3$, $4$, and $6$ influence the average amplitude. In linear diagrammatics, these events give rise to the average Green's function $\langle G({\bf r},{\bf r'})\rangle$.  However, since it
is not evident how the concept of Green's functions can be generalized to nonlinear equations, we will postpone a more detailed discussion to Sec.~\ref{sindex}.

\begin{figure}
\centerline{\includegraphics[width=8cm]{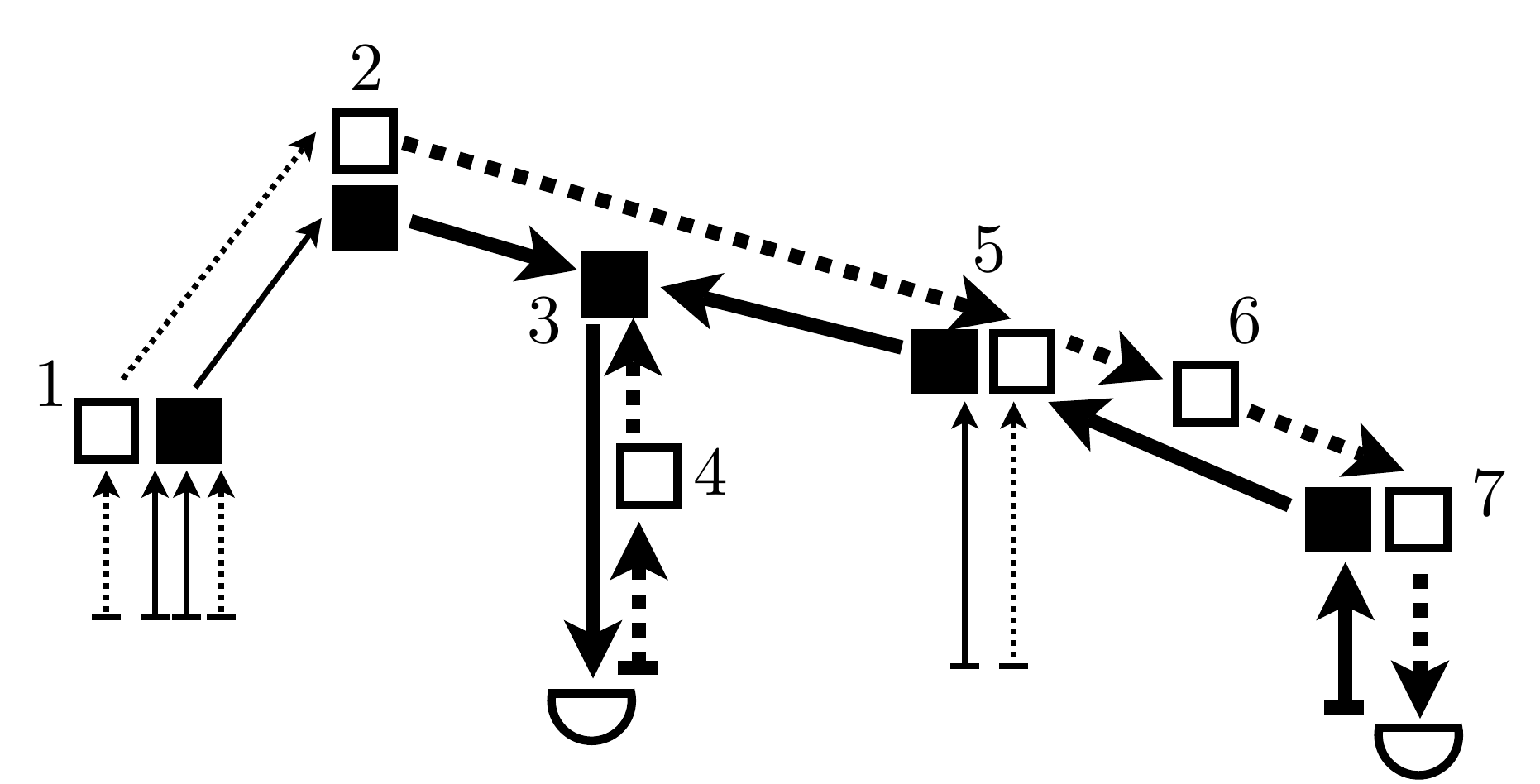}}
\caption{Example of a nonlinear crossed diagram. It is constructed in the same way as the ladder diagrams, see Fig.~\ref{ladder}, with the difference that pairs of arrows may also point in opposite directions. The reversed pairs of arrows (thick lines) define the crossed scattering path.
\label{crossed}} 
\end{figure}

In a similar way as the ladder diagrams, we can also construct crossed diagrams,
see Fig.~\ref{crossed}. Here, in contrast to the ladder diagrams, the two
finally outgoing arrows originate from two different scatterers. Between those two scatterers,
one can 
identify a
\lq crossed scattering path\rq\ by following the reversed arrow pairs. 

The mathematical expression for the diagram Fig.~\ref{crossed} reads:
\begin{eqnarray}
& & {\mathcal N}^7 g_0^3 \left(g_0^*\right)^5 g_1^2 g_1^*\int d{\bf r_1}\dots d{\bf r_7} e^{-i{\bf k_D}\cdot {\bf r_3}-i{\bf k_L}\cdot {\bf r_4}}\nonumber\\
& &  \left|G_{12}\right|^2 G_{23}G_{25}^*G_{43}^*G_{53}G_{56}^*G_{67}^*G_{75}e^{i({\bf k_L}+{\bf k_D})\cdot {\bf r_7}}\label{crossedexample}
\end{eqnarray}
Since $G_{ij}=G_{ji}$, the two expressions Eqs.~(\ref{ladderexample},\ref{crossedexample}) for the ladder and crossed diagram are identical in the exact backscattering direction ${\bf k_D}=-{\bf k_L}$.
This property is a consequence of the reciprocity symmetry. Unlike for the linear case, however, it does not imply that, in total, the crossed and ladder contributions to the backscattering signal are identical. The reason is that, in general, a nonlinear ladder diagram has more than one crossed counterpart. 
Fig.~\ref{crossed}, for example, just shows one of several different ways of reversing the scattering paths of Fig.~\ref{ladder}. Indeed, reversing the direction of arrows in all possible ways (consistent with the general structure of nonlinear field diagrams, see Sec.~\ref{snlfield}), we 
see that the outgoing solid arrow may originate either from scatterer 1, 3, or 5, and the dashed one from 1, 5, or 7.
This yields in total 2 different ladder and 7 different crossed diagrams, whose value (for  ${\bf k_D}=-{\bf k_L}$)  is identical to Eq.~(\ref{ladderexample}).
Hence, unlike to the linear case, nonlinear coherent backscattering does not originate from 
interference between only two waves propagating along reversed scattering paths,
but must be described as a multi-wave interference phenomenon.
 
For this reason, the nonlinear crossed contribution cannot simply be extracted from the ladder intensity by employing the reciprocity symmetry, but must be calculated separately. In the following two sections, we will show how the ladder and crossed diagrammatic series can be summed up into a closed form, thereby obtaining nonlinear transport equations for the ladder and crossed intensities.  
 
\section{Incoherent transport: summation of ladder diagrams}
\label{sladder}

\begin{figure}
\centerline{\includegraphics[width=8cm]{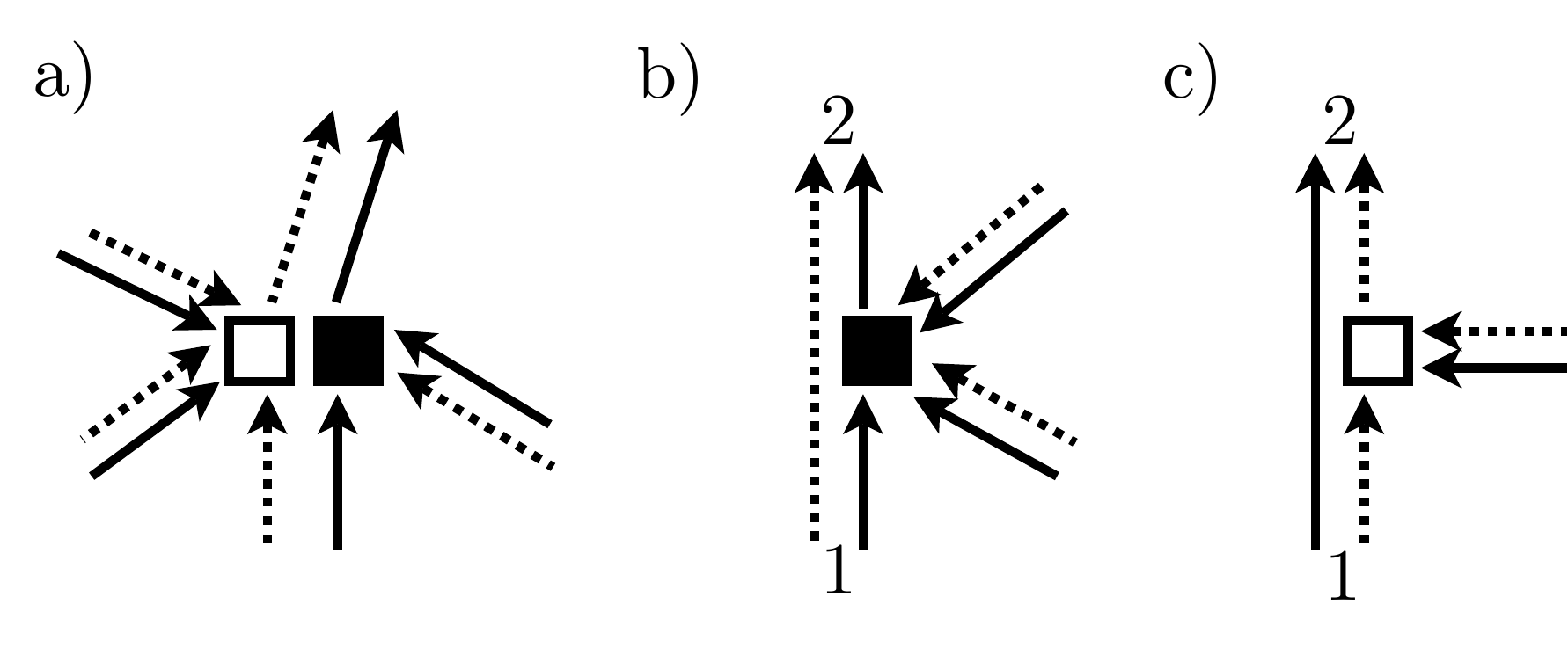}}
\caption{Building blocks for nonlinear ladder diagrams. Diagrams where $\blacksquare$ and $\Box$ occur as a pair (a), contribute to the average scattered intensity 
$K$, see Eqs.~(\ref{k1},\ref{k}), 
whereas (b) and (c) represent average propagation $\langle G\rangle$ and $\langle G^*\rangle$, see Eq.~(\ref{green}).
\label{ladderbb}} 
\end{figure}

In this section, we will express the sum of all ladder diagrams in terms of an integral equation for
the average intensity  $\langle I({\bf r})\rangle$.
For this purpose, we first identify the building blocks out of which all nonlinear ladder diagrams can
be composed. These building blocks are shown in Fig.~\ref{ladderbb}. As already mentioned in Sec.~\ref{ssladder}, we distinguish between scattering events for the average intensity represented by a pair of single events  ($\blacksquare$ and $\Box$)
 as in Fig.~\ref{ladderbb}(a), on the one hand,
and average field propagation, Fig.~\ref{ladderbb}(b,c), on the other one.

As shown below, the average propagation events modify the coherent part $I_c({\bf r})$ of the intensity,
which is defined as the square modulus of the average field amplitude, i.e.
$I_c({\bf r})=|\langle E({\bf r})\rangle|^2$. It represents the intensity originating directly from the incident field ${\bf k_L}$, and propagating until position $\bf r$ without being
scattered into another direction ${\bf k}\neq {\bf k_L}$. 
The remaining part of the average intensity, called
the diffuse intensity $I_d({\bf r})$ and defined by
\begin{equation}
\langle I({\bf r})\rangle=I_c({\bf r})+I_d({\bf r})
\end{equation}
describes the part of the average intensity which has undergone at least one scattering event. 

\subsection{Nonlinear scattering and diffuse intensity}

Thus, the diffuse intensity at point ${\bf r}$ inside the sample volume is obtained as the sum of the intensities radiated by each of the individual scatterers at positions ${\bf r'}$:
\begin{equation}
I_d({\bf r})=\int_V d{\bf r'} \left|\langle G({\bf r},{\bf
  r'})\rangle\right|^2
K({\bf r'})\label{id}
\end{equation}
where $K({\bf r'})$ describes the average intensity  emitted from point ${\bf r'}$, and 
$|\langle G({\bf r},{\bf r'})\rangle|^2$ the subsequent propagation of the intensity from 
${\bf r'}$ to $\bf r$.

\subsubsection{Average scattered intensity}
\label{sscatt}

We first focus our attention on the average scattered intensity $K({\bf r'})$, which we now calculate by means of 
the nonlinear scattering diagrams, Fig.~\ref{ladderbb}(a).

The example shown in Fig.~\ref{ladderbb}(a) shows a pair of scattering events $f$ ($\blacksquare$) with nonlinear order $n=1$, and $f^*$ ($\Box$) with nonlinear order $m=2$ (see Sec.~\ref{sssingle}).
To obtain the general result, we have to take the sum over $n$ and $m$, respectively.
Hence, the
total number of incoming solid and dashed arrows in Fig.~\ref{ladderbb}(a) equals $n+m+1$.
According to the general recipe for constructing ladder diagrams (see Sec.~\ref{ssladder}),
all arrows are grouped in pairs of solid and dashed arrows, respectively.
Among these $n+m+1$ arrow pairs, some represent the diffuse intensity $I_d$, and the
remaining ones the coherent intensity $I_c$. 
If we now label  all incident arrows with a certain index,
for example $E_0$ for the coherent mode, and $E_i$, with $i>0$, for a wave emitted by scatterer $i$,
summing over all incident fields amounts to calculating the average value of $I^{n+m+1}$ as follows:
\begin{eqnarray}
\langle I^{n+m+1}\rangle & = & \left<\left(\sum_{i,j}E_iE_j^*\right)^{n+m+1}\right>\label{avint1}\\
& \simeq &  \sum_{k=0}^{n+m+1} {n+m+1 \choose k}^2 k! I_d^k
  I_c^{n+m+1-k}\ \ \ \ \ \  \label{avint2}
  \end{eqnarray}
  with $I_c=|E_0|^2$ and $I_d=\sum_{i>0} |E_i|^2$ (see below).
 Within the ladder approximation, only such terms contribute in average to  Eq.~(\ref{avint1}),
 where each amplitude $E_i$ is accompanied by its complex conjugate $E_i^*$.
 Due to the large number of scatterers, we may 
 furthermore neglect the cases where more than one arrow pair $|E_i|^2$, $i\neq 0$, is emitted by the same scatterer. This leads to Eq.~(\ref{avint2}), which can be interpreted as follows:
 among all $n+m+1$ incoming arrow pairs,
 we select $k$ solid and $k$ dashed arrows contributing to $I_d$, and take into 
 account a factor $k!$ for all possible pairings between them. 
 
 Furthermore, note that we have identified in Eq.~(\ref{avint2}) the sum $\sum_{i>0} |E_i|^2$
 of the intensities radiated by each scatterer with the diffuse intensity $I_d$ given by Eq.~(\ref{id}).
 This is justified since, when forming a ladder diagram by
 connecting several building blocks like the one shown in Fig.~\ref{ladderbb}(a) to each other,
 every outgoing arrow pair serves as an incoming arrow pair for the next building block.
 
  Finally, the average scattered intensity results as:
\begin{equation}
K(I_d,I_c)={\mathcal N}\sum_{n,m=0}^\infty g_n g_m^*
\left< I^{n+m+1}\right>\label{k1}
\end{equation}
with ${\mathcal N}=N/V$ the density of scatterers, and
$\left< I^{n+m+1}\right>$ given by Eq.~(\ref{avint2}).
Since $I_d$ and $I_c$ depend on ${\bf r}$, also $K$ is a function of $\bf r$, as written in Eq.~(\ref{id}). 
Hence, Eq.~(\ref{id}) forms a self-consistent
equation for the diffuse intensity $I_d({\bf r})$. Solving this integral equation by iteration,
we explicitly obtain the sum of all ladder diagrams, i.e. all possible connections of the building blocks
shown in Fig.~\ref{ladderbb}.

\subsubsection{Statistical properties of the wave intensity}
\label{ssstat}

The statistical distribution of the wave intensity, which is expressed in terms of its statistical moments 
in Eq.~(\ref{avint2}),
can also be derived in an
alternative way as follows: we assume that real and imaginary part of the 
diffuse field $E_d$, defined by $E_d=E-\langle E\rangle$, follow
a Gaussian distribution with mean value zero and standard
deviation $\sqrt{I_d/2}$. This
so-called ``chaotic'' distribution for the diffuse field results from
a random-phase approximation, i.e. by adding many partial waves 
(i.e. $E_d=\sum_{i>0}E_i$)
with
independent random phases, thereby neglecting any correlations between them \cite{goodman}. 
In this sense, the calculation of the average intensity by means
of ladder diagrams is equivalent to neglecting correlations between
different scatterers. This approximation is known to be well
fulfilled in the case of a dilute medium ($k\ell\gg 1$). 

The above considerations allow us to rewrite Eq.~(\ref{k1}) in a more compact form:
\begin{equation}
K={\mathcal N}\left<|f|^2\right>={\mathcal N}\left<I|g(I)|^2\right>\label{k}
\end{equation}
where the average is taken with respect to the chaotic distribution
for the diffuse field. Note that the result of this average depends only on $I_d$ and
$I_c=|\langle E\rangle|^2$. For the explicit calculation of Eq.~(\ref{k}), it is useful to
know the corresponding distribution function of the wave intensity $I$:
\begin{equation}
P(I)=\frac{e^{-(I+I_c)/I_d}}{I_d} I_0\left(2\sqrt{\frac{I
    I_c}{I_d^2}}\right),\label{pi}
\end{equation}
where $I_0$ denotes the modified Bessel function of the first kind.
In the absence of coherent intensity (i.e. deep inside the bulk), Eq.~(\ref{pi}) reduces to the
Rayleigh distribution $P(I)|_{I_c=0}=\exp(-I/I_d)/I_d$. On the other hand, for vanishing diffuse intensity
(i.e., a scattering medium with extremely small optical thickness),
the fluctuations vanish, such that $P(I)|_{I_d\to 0}\to \delta(I-I_c)$.

Using the intensity distribution function, Eq.~(\ref{pi}), the average scattered intensity, Eq.~(\ref{k}), results as:
$K(I_d,I_c)={\mathcal N}\int dI P(I) I |g(I)|^2$.
In the following, we will also need averages of the form $\langle E h(I)\rangle$, where 
$h(I)$ may be an arbitrary function of $I$. Here, the  relation
\begin{equation}
\langle E h(I)\rangle=\frac{\langle
Ih(I)\rangle-I_d\langle h(I)+Ih'(I)\rangle}{\langle E^*\rangle}\label{pi2}
\end{equation}
can be shown to result from the Gaussian diffuse field distribution.

\subsection{Nonlinear average propagation and coherent intensity}
\label{sindex}

In order to complete the transport equation
(\ref{id}), the coherent intensity
$I_c=|\left<E\right>|^2$ and the average Green's
function $\langle G({\bf r_1},{\bf r_2})\rangle$ remain to be calculated. Both result from the
diagram shown in Fig.~\ref{ladderbb}(b).

\subsubsection{Average Green's function and nonlinear refractive index for the diffuse intensity} 

This diagram can be interpreted as follows: the scattering event ($\blacksquare$) modifies the propagation $G_{12}$ represented by the solid arrows from 1 to 2. Integrating over the position $\bf r_3$ of the scattering event, it
turns out that, if $kr_{12}\gg 1$ (dilute medium),
only positions situated on the straight line connecting
${\bf r_1}$ and ${\bf r_2}$ contribute to the integral. (In other words: average propagation results from
forward scattering.) The phase factors relevant for this stationary phase argument originate from the two vacuum Green's functions $G_{13}$ and $G_{32}$, i.e. the two vertical solid arrows pointing from the starting point $1$ and to the end point $2$, respectively.
In leading
order of $1/(kr_{12})$, this integral yields:
\begin{equation}
{\mathcal N}\int d{\bf r_3} G_{13}G_{32}\simeq
\frac{\mathcal N}{2k^2} ikr_{12} G_{12}
\label{1storder}
\end{equation}
The result on the right hand side should be compared with the
expression $ikr_{12} (n-1) G_{12}$, which is obtained if one inserts a
refractive index $n$ in the exponent of the vacuum Green's function,
Eq.~(\ref{green0}), and expands the resulting expression in first
order of $n-1$.  Hence, we can derive an expression
for the refractive index $n$ as follows:
we sum over all possibilities to group the
incoming arrows into pairs. Since there are $m+1$ incoming solid and $m$ dashed arrows,
we obtain:
\begin{equation}
n-1 = \frac{\mathcal N}{2k^2}\sum_{m=0}^\infty g_m (m+1)\langle I^m\rangle
=\frac{\mathcal N}{2k^2}
\left<\frac{\partial f}{\partial E}\right>\label{index}
\end{equation}
with $\langle I^m\rangle$ given by Eq.~(\ref{avint2}).
The second equality results from taking the partial 
derivative of $f$, see Eq.~(\ref{gn}), with respect to $E$ (where
$E$ and $E^*$ are treated as independent variables). 
The imaginary part of the refractive index defines the extinction path length $\ell$ according to:
\begin{equation}
\frac{1}{\ell}=2k{\rm Im}\{n\} \label{ell}
\end{equation}
For a conservative medium, i.e. for nonlinear scatterers of the form given by Eq.~(\ref{gcons}),
the extinction path length is equal to the mean free path.

Since the intensity distribution function $P(I)$, Eq.~(\ref{pi}), with respect to which the mean value
$\langle\dots\rangle$ in Eq.~(\ref{index}) is evaluated, depends on $I_c({\bf r})$ and $I_d({\bf r})$, the
nonlinear refractive index attains a
spatial dependence $n({\bf r})$. To obtain  the average Green's function
describing propagation between two positions ${\bf r}_1$ and ${\bf r}_2$,
we now integrate the refractive index along the optical path (i.e. points of stationary phase)
connecting ${\bf  r}_1$ and ${\bf r}_2$, which, as mentioned in Sec.~\ref{setupsa}, we approximate by
a straight line:
\begin{equation}
\langle G({\bf r}_1,{\bf r}_2)\rangle=\frac{\exp\left(ik|{\bf r}_1-{\bf
  r}_2|\int_0^1 ds n({\bf r}_s)\right)}{4\pi|{\bf r}_1-{\bf r}_2|}\label{green}
\end{equation}
with ${\bf r}_s=s{\bf r}_1+(1-s){\bf r}_2$. By writing $n$ in the exponent, we effectively sum up a series of  diagrams with not only one, as in Fig.~\ref{ladderbb}(b), but an arbitrary number of forward scattering events occurring between $\bf r$ and $\bf r'$. 

From Eq.~(\ref{green}), we obtain $|\langle G({\bf r}_1,{\bf r}_2)\rangle|^2$, which appears in the transport equation (\ref{id}), as
\begin{equation}
|\langle G({\bf r}_1,{\bf r}_2)\rangle|^2=\frac{\exp\left(-|{\bf r}_1-{\bf
  r}_2|\int_0^1 ds/ \ell({\bf r}_s)\right)}{(4\pi|{\bf r}_1-{\bf r}_2|)^2}\label{green2}
\end{equation}

The name "average Green's function" must be taken with some care, since, strictly speaking, nonlinear equations like Eq.~(\ref{model1}) do not have a well defined Green's function. However, a Green's function can be defined for a linearized version of the wave equation (\ref{model1}).
The polarizability of a single scatterer in response to a small probe field $\Delta E$ is then
proportional to $\langle \partial f/\partial E\rangle$, which yields the refractive index, Eq.~(\ref{index}), and corresponding average Green's function, Eq.~(\ref{green}). The probe field interpretation is also consistent with the observation that, due to the diluteness of the medium,
the field emitted by any individual scatterer is very weak compared to the incident field or the sum of the 
fields emitted by all other scatterers.
In the following, we will hence continue speaking of "average Green's functions" when referring to the average propagation between two scattering events.

\subsubsection{Refractive index for the coherent intensity}

The probe field argument, however, does not apply for the coherent field $\langle E({\bf r})\rangle$, which is not a weak field - at least close to the boundary of the medium, where it equals the incident field. In the diagrammatic calculation of Fig.~\ref{ladderbb}(b), this difference is reflected by the fact that the vertical solid arrow starting from point 1 in Fig.~\ref{ladderbb}(b) now originates from the coherent instead from a diffusive mode. This renders the combinatorial counting slightly different:
we select $l$ out of the $m$ incoming dashed
arrows and $l$ out of the $m+1$ solid arrows for the diffuse modes,
and group them in pairs in $l!$ possible ways. Thereby, we obtain 
the refractive index for the coherent mode as follows:
\begin{eqnarray}
n_c-1 & = & \frac{\mathcal N}{2k^2}\sum_{m=0}^\infty g_m \sum_{l=0}^m
{m+1 \choose l}{m
  \choose l} l! I_d^l I_c^{m-l}\ \ \ \ \label{nc1}\\
& = & \frac{\mathcal N}{2k^2}
\frac{\langle f\rangle}{\langle E\rangle}\label{nc}
\end{eqnarray}
and $1/\ell_c=2k{\rm Im}\{n_c\}$.
Eq.~(\ref{nc}) is derived from Eq.~(\ref{nc1}) with help of Eq.~(\ref{pi2}). 

The fact that the refractive indices are different for diffuse and
coherent waves should not be surprising:
because of the nonlinearity, the effective medium
is modified by the
propagating waves themselves, and, due to their different statistical
properties, this is done in a different way by coherent and diffusive
waves, respectively. From nonlinear optics in Kerr media, for example, it is known \cite{boyd} that
a strong wave affects the refractive index of a weak wave of the same frequency twice as much
as it affects its own refractive index, cf. Eqs.~(\ref{nhom},\ref{nchom}) below.

Using the refractive index $n_c({\bf r})$, the coherent field results as:
\begin{equation}
\langle E({\bf r})\rangle  =  \exp\left(ik \int_0^z dz' n_c({\bf r}_{z'})\right)\label{ec}
\end{equation}
where $z$ denotes the
distance from the surface of $V$ to $\bf r$, in the direction  
of the incident beam, and ${\bf r}_{z'}={\bf r}-z'{\bf k_L}/k$. The corresponding 
coherent intensity reads:
\begin{equation}
I_c({\bf r})=\left|\langle E({\bf r})\rangle\right|^2=\exp\left(- \int_0^z dz' /\ell_c({\bf r}_{z'})\right)\label{ic}
\end{equation}
Finally, let us remark that the above treatment of average propagation must be modified in the case where the incident wave is a superposition of several plane waves.  In this case, additional coherent components are produced by four-wave mixing processes \cite{boyd}. These are described by diagrams which are neither ladder nor crossed diagrams, but nevertheless fulfill a phase-matching condition, like for example $\exp[i({\bf k_1}+{\bf k_2})\cdot{\bf r}]=1$ in the case ${\bf k_1}=-{\bf k_2}$ of two opposite incident waves.

\subsection{Incoherent component of the backscattered intensity}

We have now all ingredients at hand to determine the finally measured quantity, the intensity 
$\Gamma({\bf k_D})$ scattered into direction ${\bf k_D}$, see Eq.~(\ref{gamma}).
First, we solve the coupled system of integral equations, Eqs.~ (\ref{id},\ref{ic}), in order to find
the coherent and diffuse intensities $I_c({\bf r})$ and $I_d({\bf r})$. These, in turn, determine the intensity $K({\bf r})$ emitted from ${\bf r}$, see Eq.~(\ref{k}), the refractive index $n({\bf r})$, see Eq.~(\ref{index}), and the extinction path length $\ell({\bf r})$, see Eq.~(\ref{ell}).
Finally, the bistatic ladder coefficient results as:
\begin{equation}
\Gamma_L({\bf k_D})  = \int \frac{d{\bf r}}{4\pi A}
\exp\left(-\int_0^{z_{\bf q}}dz'/\ell({\bf r}_{z'})\right) K({\bf r})\label{gammal}
\end{equation}
where $z_{\bf q}$ denotes the distance from $\bf r$ to the surface of
$V$ in the direction $\bf k_D$ of the outgoing beam, and
${\bf r}_{z'}={\bf r}+z'{\bf k_D}/k$.

\section{Coherent transport: summation of crossed diagrams}
\label{scrossed}

The bistatic ladder coefficient, Eq.~(\ref{gammal}), expresses the incoherent part of the backscattered intensity, where all interferences are neglected. For a dilute medium, this is a good approximation if
the detector is not placed in exact backscattering direction ($|{\bf k_D}+{\bf k_L}|>1/\ell$), but for
${\bf k_D}=-{\bf k_L}$, the coherent backscattering contribution resulting from the crossed diagrams must be taken into account.

\subsection{Building blocks for crossed diagrams}

\begin{figure}
\centerline{\includegraphics[width=8cm]{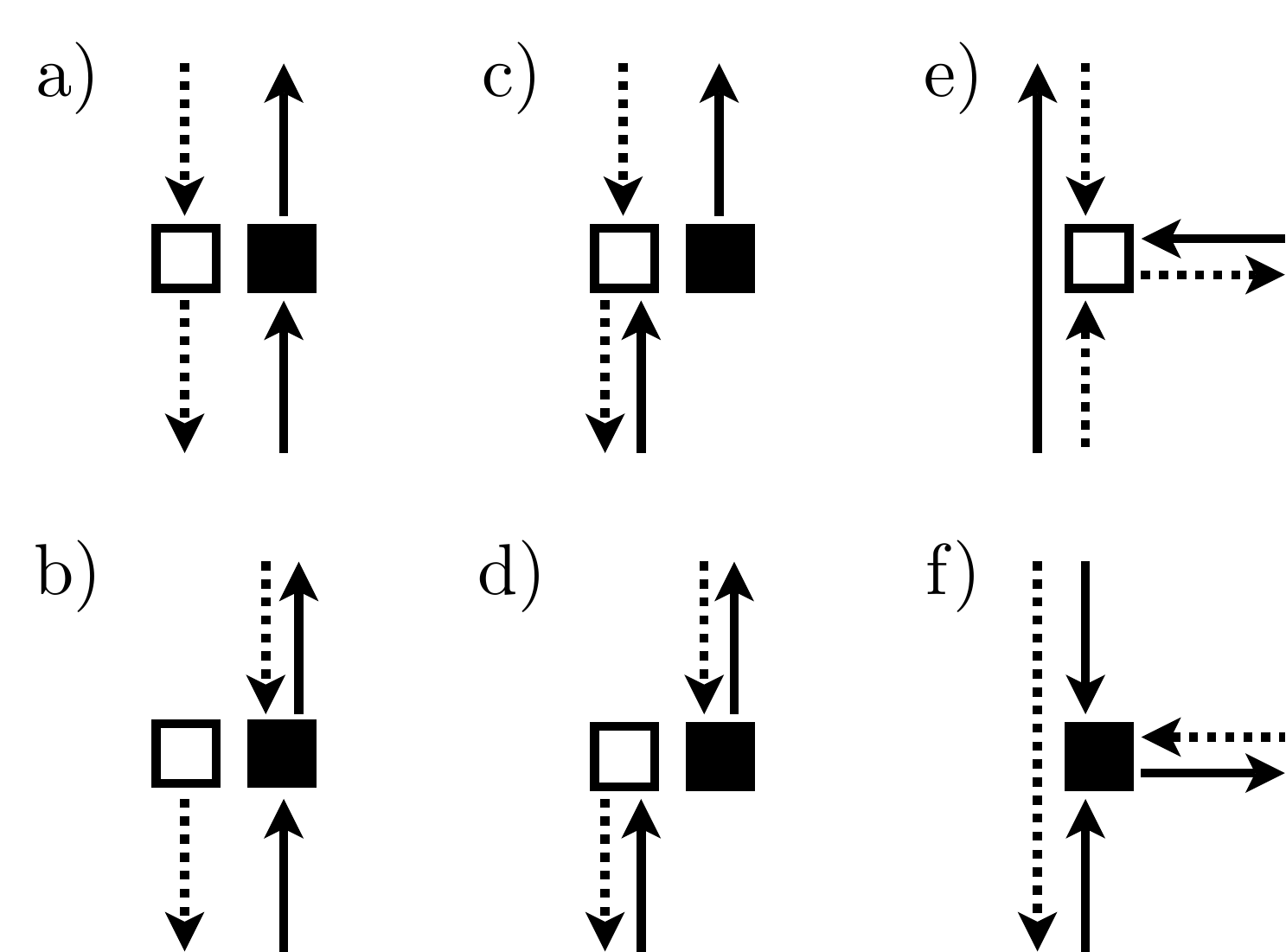}}
\caption{Building blocks for crossed diagrams. By connecting these building blocks, we can construct any crossed scattering path, see the thick lines in Fig.~\ref{crossed}. Diagrams (a-d) represent the crossed scattering events $\kappa^{(a-d)}$, see Eqs.~(\ref{kappaa}-\ref{kappad}), and (e,f) the crossed  propagation events $\tau$, see Eq.~(\ref{tau}), and $\tau^*$, respectively.
\label{crossedbb}} 
\end{figure}

In analogy to the summation of ladder diagrams in Sec.~\ref{sladder}, we first identify the individual building blocks
out of which any crossed diagram is composed. These are shown in Fig.~\ref{crossedbb}. For simplicity, we only display the reversed arrow pairs forming the crossed scattering path, which are drawn as thick lines in the example of
Fig.~\ref{crossed}.  It is sufficient to concentrate on the crossed path, since
the remaining elements, i.e., the thin lines in Fig.~\ref{crossed}, all form ladder diagrams, and hence can be expressed in terms of the coherent and diffuse intensities $I_{c,d}({\bf r})$, according to the results of the previous section.

\subsubsection{Crossed scattering diagrams}

Like for the ladder building blocks, Fig.~\ref{ladderbb}, we distinguish between scattering and propagation diagrams. The crossed scattering diagrams,
Fig.~\ref{crossedbb}(a-d) are very similar to the ladder scattering
diagram, Fig.~\ref{ladderbb}(a). The only difference is that, in the ladder case, the two outgoing arrows
are always grouped together, whereas in the crossed case, the outgoing arrows are grouped together with incoming arrows. According to whether or not the incoming arrows are connected to the same scattering event as the corresponding outgoing arrow, we obtain the four possibilities displayed in Fig.~\ref{crossedbb}(a-d). Remember  that, as mentioned above,  the crossed diagrams may be supplemented by an arbitrary number of incoming ladder pairs (thin lines in Fig.~\ref{crossed}). In particular, in the case of Fig.~\ref{crossed}(b,c) and (d), at least one incoming ladder pair is required in order to obtain  diagrams consistent with the general structure described in Sec.~\ref{snlfield}, according to which each nonlinear event exhibits one outgoing arrow and $n+1$ or $n$ incoming solid or dashed arrows. 

For the explicit calculation of  diagrams Fig.~\ref{crossedbb}(a-d), we first denote, as in the
calculation of diagram Fig.~\ref{ladderbb}(a), see Eq.~(\ref{avint2}), the
number of dashed arrows entering the filled square ($\blacksquare$) by $n$ (which
determines the number of solid arrows entering this square as $n+1$),
and, likewise, the number of solid (or dashed) arrows entering the open
square ($\Box$) by $m$ (or $m+1$). Then, one of the incoming solid arrows and one of the incoming dashed arrows must be chosen as partner for the outgoing dashed and solid arrow, respectively. According to which of the  cases (a-d) we consider, this yields a factor $n+1$ or $m$ for the solid arrows, and $m+1$ or $n$ for the dashed ones. After that, 
the number of remaining arrow pairs is decreased by one, i.e., $n+m$ instead of $n+m+1$ in
Eq.~(\ref{k1}). Thereby, we obtain:
\begin{eqnarray}
\kappa^{(a)} & = & \mathcal{N}\sum_{m,n}g_ng_m^*(n+1)(m+1)\langle I^{m+n}\rangle\\
\kappa^{(b)} & = & \mathcal{N}\sum_{m,n}g_ng_m^*(n+1)n\langle I^{m+n}\rangle\\
\kappa^{(c)} & = & \mathcal{N}\sum_{m,n}g_ng_m^*(m+1)m\langle I^{m+n}\rangle\\
\kappa^{(d)} & = & \mathcal{N}\sum_{m,n}g_ng_m^*nm\langle I^{m+n}\rangle
\end{eqnarray}
which, using Eq.~(\ref{gn}), we rewrite as:
\begin{eqnarray}
\kappa^{(a)} 
& = & {\mathcal N}\left<\frac{{\partial }f}{{\partial }E}\frac{{\partial }f^*}{{\partial }E^*}\right>\label{kappaa}\\
\kappa^{(b)} 
& = & {\mathcal N}\left<\frac{{\partial }f}{{\partial }E}\frac{{\partial }f}{{\partial }E^*}\right>\label{kappab}\\
\kappa^{(c)} 
& = & {\mathcal N}\left<\frac{{\partial }f^*}{{\partial }E}\frac{{\partial }f^*}{{\partial }E^*}\right>\label{kappacc}\\
\kappa^{(d)} 
& = & {\mathcal N}\left<\frac{{\partial }f^*}{{\partial }E}\frac{{\partial }f}{{\partial }E^*}\right>\label{kappad}
\end{eqnarray}
Note that $\kappa^{(c)}=\left(\kappa^{(b)}\right)^*$, whereas $\kappa^{(a)}$ and $\kappa^{(d)}$ are real.
Let us remind again that the mean values $\langle\dots\rangle$ are taken  
with respect to the intensity distribution function discussed in Sec.~\ref{ssstat}, see Eq.~(\ref{pi}).

As already discussed above in the case of average propagation, see Sec.~\ref{sindex}, the expressions on the right hand side of Eqs.~(\ref{kappaa}-\ref{kappad}) can be interpreted in terms of probe fields. Here, we need two independent probe fields $\Delta E$ and $\Delta E^*$, which represent the two reversed fields propagating along the crossed scattering path. The probe fields act as derivatives ${\partial }/{\partial }E$ or ${\partial }/{\partial }E^*$ 
either on $f$ or $f^*$, what yields in total four different cases (a-d). Note that in the case of linear scatterers ($f=gE$, with $g={\rm const.}$), cases (b-d) vanish, and only case (a) survives.

To simplify the notation in the following equations, we define:
\begin{eqnarray}
\kappa & = & \kappa^{(a)}+\kappa^{(c)}=\mathcal{N}\left<\frac{\partial }{{\partial }E}\left(f\frac{{\partial }f^*}{{\partial }E^*}\right)\right>\label{kappa}\\
\tilde{\kappa} & = &
\kappa^{(b)}+\kappa^{(d)}=\mathcal{N}\left<\frac{\partial }{{\partial }E}\left(f^*\frac{{\partial }f}{{\partial }E^*}\right)\right>\label{tildekappa}
\end{eqnarray}
We note that $\kappa-\tilde{\kappa}^*\in\mathbb R$.

If one of the incoming arrows participating in the crossed path
originates from the coherent mode, the above calculation is modified
in a similar way as for Eq.~(\ref{nc1}). As in Eq.~(\ref{nc}), the
result is that, in the case of a solid arrow, 
the derivative $\partial /({\partial }E)$ is replaced by $1/\langle
E\rangle$:
\begin{equation}
\kappa_c  = 
\frac{\mathcal{N}}{\langle E\rangle}\left<f\frac{{\partial }f^*}{{\partial }E^*}\right>,\ \ 
\tilde{\kappa}_c  = 
\frac{\mathcal{N}}{\langle E\rangle}\left<f^*\frac{{\partial }f}{{\partial }E^*}\right>\label{kappac}
\end{equation}
If the dashed arrow originates from the coherent mode, the complex
conjugate expressions $\kappa_c^*$ and $\tilde{\kappa}_c^*$ must be
used. The case where both solid and dashed arrows originate from the
coherent mode can be excluded, since the resulting diagram would be
identical to a ladder diagram (single scattering).  

\subsubsection{Crossed propagation diagrams}

Concerning nonlinear propagation, we can use the same average Green's function, Eq.~(\ref{green}), and refractive indices, 
Eqs.~(\ref{index},\ref{nc}) as in the ladder case, since average propagation is an amplitude property, and symmetric against reversing the direction of propagation.
To demonstrate this equality on the diagrammatic level, one just reverses the direction of the
vertical ``spectator arrow'' between 1 and 2 in diagrams
Fig.~\ref{ladderbb}(b,c).
However, it is also possible to reverse the
direction of one of the arrows contributing to the 
``pump intensity'', i.e. the arrows coming from the left side in
Fig.~\ref{ladderbb}(b,c). This yields two new diagrams, which are
displayed in Fig.~\ref{crossedbb}(e,f).

The physical interpretation of these diagrams is not straightforward:
on one hand, they are very similar to the nonlinear propagation diagrams,
Fig.~\ref{ladderbb}(b,c), from which they result just by exchanging two arrows.
On the other hand, they can also be interpreted as 
nonlinear scattering diagrams, since, as clearly evident from
Fig.~\ref{crossedbb}(e,f), the crossed path changes its direction of
propagation due to these events. 
Their mathematical evaluation, however,
poses no difficulties. From Fig.~\ref{crossedbb}(e), we see that we
have to select two dashed and one solid arrow from all the arrows
entering the nonlinear event. If the total number of incoming solid (or
dashed) arrows is denoted by $m$ (or $m+1$), we obtain:
\begin{eqnarray}
\tau  & = &  -\frac{i\mathcal  N}{2k}\sum_{m=1}^\infty g_m^* (m+1)m^2\langle I^{m-1}\rangle\\
&  =  & -i\frac{\mathcal N}{2k}
\left<\frac{\partial ^3f^*}{({\partial }E^*)^2{\partial }E}\right>\label{tau}
\end{eqnarray}
Diagram Fig.~\ref{crossedbb}(f) just yields the complex conjugate $\tau^*$.
The prefactor $-i/(2k)$ in Eq.~(\ref{tau}) results from the integration over the position of the nonlinear event, see Eq.~(\ref{1storder}). If one of the incident fields originates from the coherent mode, ${\partial }/{\partial }E$ (or ${\partial }/{\partial }E^*$) is again replaced by $1/\langle E\rangle$ (or $1/\langle E^*\rangle$), i.e. 
\begin{equation}
\tau_c=-\frac{i{\mathcal N}}{2k\langle E\rangle}\left<\frac{\partial^2f^*}{({\partial }E^*)^2}\right>=-\frac{i{\mathcal N}}{2k\langle E^*\rangle}\left<\frac{\partial^2f^*}{{\partial }E{\partial }E^*}\right>\label{tauc}
\end{equation} 
Note also the following difference between the crossed scattering and propagation diagrams, Figs.~\ref{crossedbb}(a-d) and (e,f), respectively:
whereas in (a-d), only reversed arrow pairs are explicitly displayed, (e,f) also contain a ladder intensity entering the diagram from  below (e) or above (f). In the exact version of the transport equations, see Eqs.~(\ref{ccex}-\ref{coex}), these will be represented
by $K({\bf r'})$, i.e. the ladder intensity emitted from ${\bf r'}$.  In general, ${\bf r'}$ differs from the position ${\bf r}$ where the  nonlinear event $\tau({\bf r})$ occurs, but if we assume that $K({\bf r'})$ varies slowly on the length scale $\ell$, we may set ${\bf r'}={\bf r}$. In this approximation, the contribution of the diagrams (e,f) to the cooperon cross sections will be given by $\tau({\bf r})K({\bf r})\ell({\bf r})$ and $\tau^*({\bf r})K({\bf r})\ell({\bf r})$, see Eqs.~(\ref{sigma},\ref{tildesigma}) below.

\subsection{Connecting the building blocks}
\label{sforbidden}

\subsubsection{Forbidden diagrams}
\label{sforbidden}
\begin{figure}
\centerline{\includegraphics[width=8cm]{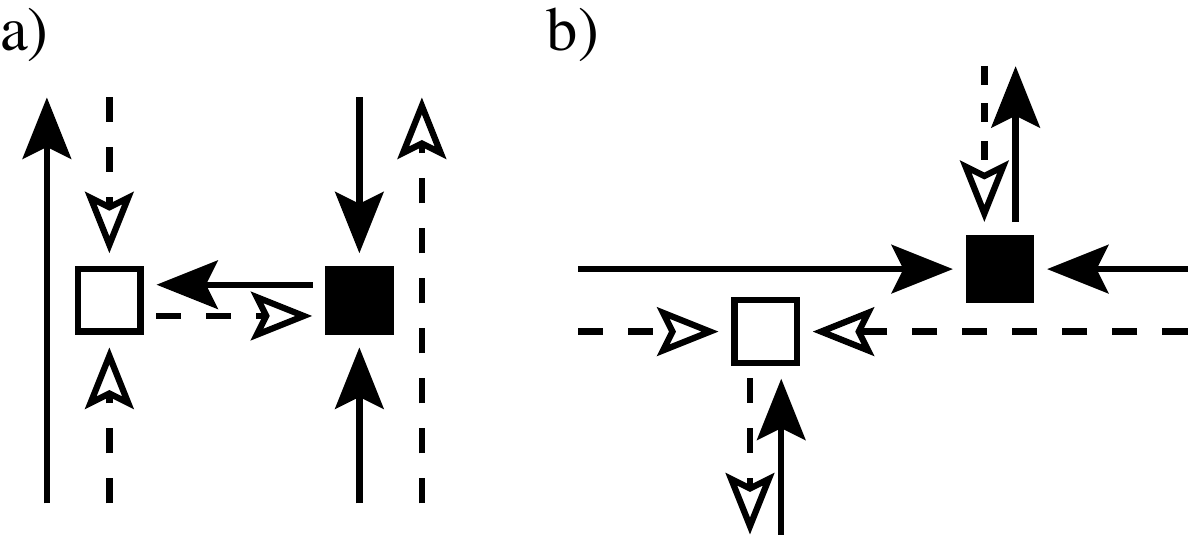}}
\caption{Examples of forbidden (a) and allowed (b) combinations of the
  building blocks shown in Fig.~\ref{crossedbb}. The forbidden combinations
  correspond to those forming a closed loop where two nonlinear events are
  connected by two Green's functions pointing in
  different directions.
\label{forbidden}} 
\end{figure}

After having discussed the building blocks shown in
Fig.~\ref{crossedbb}, we now examine how they can be connected with
each other in order to form a crossed scattering path. Like in the case of the ladder
diagrams, see Sec.~\ref{sladder}, our aim is to derive an integral
equation describing propagation along the crossed scattering path
through the scattering volume. We call the quantity which is
transported along this path the ``crossed
intensity'' $C({\bf r})$. To define the direction of propagation, we will choose the solid arrows in the following.  Accordingly, we
identify in each of the building blocks of Fig.~\ref{crossedbb}
an ``incoming'' and ``outgoing'' crossed intensity according to the direction of the solid arrow.

Similarly to the ladder case, see Sec.~\ref{sladder}, we could now
obtain a crossed transport equation simply by identifying the
incoming and outgoing crossed intensities, representing both by the
same function $C({\bf r})$. In this way, we would build crossed
scattering paths by combining the individual building blocks of
Fig.~\ref{crossedbb} in all possible ways: each outgoing crossed
intensity may serve as an incoming crossed intensity for the next diagram.  

However, it turns out that some of the diagrams formed in this way exhibit closed loops. An example is shown in Fig.~\ref{forbidden}(a), where two scattering events ($\Box$ and $\blacksquare$) are connected by two Green's functions pointing in different directions. Such closed loops never occur in the series of nonlinear field diagrams discussed in Sec.~\ref{snlfield}.
Due to their construction as iterative solution of Eq.~(\ref{model1}), each field incident on any given scattering event must form a valid nonlinear field diagram on its own.  This property is obviously violated in Fig.~\ref{forbidden}(a), where none of the two scattering events can exist without the other one. Hence, we must avoid the occurrence of closed loops when connecting the crossed building blocks with each other.

As explained in the following, this is possible by splitting the crossed intensity $C({\bf r})$ into two parts $C_1({\bf r})$ and $C_2({\bf r})$. First, let us look at diagrams Fig.~\ref{crossedbb}(a,c) and (e). Since these are the diagrams, where the outgoing crossed intensity is emitted by two different scattering events ($\blacksquare$ and $\Box$, respectively), we can connect them in all possible ways without forming any closed scattering path. Thereby, we obtain the transport equation for $C_1({\bf r})$, see Eq.~(\ref{c1}) in the following section. If one of the events (b,d) or (f) occurs, the crossed intensity changes from type $C_1$ to $C_2$, thereby indicating that the outgoing crossed intensity is now connected to a single scattering event ($\blacksquare$). A closed scattering path would be formed if $C_2$ was connected to a diagram where the incoming crossed intensity points towards the same event ($\Box$). This is the case for (c,d) and (e). Hence, the propagation of $C_2({\bf r})$ may occur only by the remaining diagrams (a,b) and (f). This yields the transport equation for $C_2({\bf r})$, see Eq.~(\ref{c2}).

\subsubsection{Crossed transport equations}
\label{crossedtrans}

As discussed above, the forbidden diagrams can be excluded by splitting the crossed intensity into two parts  $C_1({\bf r})$ and $C_2({\bf r})$, respectively. The transport of $C_1({\bf r})$ is described by diagrams Fig.~\ref{crossedbb}(a,c) and (e), respectively. The quantities corresponding to diagrams (a) and (c) are $\kappa^{(a)}$ and $\kappa^{(c)}$, whose sum is defined as $\kappa$, see Eq.~(\ref{kappa}), 
whereas diagram (e) is approximately described by $\tau K \ell$, see Eq.~(\ref{tau}) and the discussion after Eq.~(\ref{tauc}). Thereby, the three diagrams (a,c) and (e) governing
the transport of $C_1$ are summarized by the quantity 
\begin{equation}
\sigma({\bf r})=\kappa({\bf r})+\tau({\bf r}) K({\bf r})\ell({\bf r})
\label{sigma}
\end{equation}
which we call ``cooperon cross section'' in the following. 
(More precisely, $\sigma$ corresponds to a  cross section multiplied by $4\pi\mathcal N$, with $\mathcal N$ the density of scatterers. Thus, the dimension of $\sigma$ is one over length instead of length squared.)  
The transport of $C_2$, performed by diagrams
Fig.~\ref{crossedbb}(a,b) and (f), is described by the complex conjugate
$\sigma^*({\bf r})$ of Eq.~(\ref{sigma}). 

Furthermore, the transition from
$C_1$ to $C_2$ is determined by diagrams Fig.~\ref{crossedbb}(b,d) and
(f), or
\begin{equation}
\tilde{\sigma}({\bf r})=\tilde{\kappa}({\bf r})
+\tau^*({\bf r}) K({\bf r})\ell({\bf r})
\label{tildesigma}
\end{equation}
Note that $\tilde{\sigma}=0$ in the linear case.
If the incoming or outgoing crossed intensity originates from the
coherent mode, the above expressions are modified to
$\sigma_c=\kappa_c+\tau_c K \ell$ and
      $\tilde{\sigma}_c=\tilde{\kappa}_c+\tau_c^* K \ell$.
Finally, the "crossed coherent mode" $C_c({\bf r})$ is different from $I_c({\bf r})$, Eq.~(\ref{ic}), since
only one of the two arrows (in our convention: the solid one) originates from the coherent mode
(index $n_c$), whereas the other one corresponds to an outgoing wave (index $n$).
In total, we arrive at:
\begin{eqnarray}
\label{set1}
C_c({\bf r}) & = & \exp\left(ik\int_0^z dz'
\Bigl(n_c({\bf r}_{z'})-n^*({\bf r}_{z'})\Bigr)\right)\label{cc}\\
C_{1}({\bf r}) & = & \int_V d{\bf
  r'}|\langle G({\bf r},{\bf r'})\rangle|^2\times\nonumber\\
& & \times \Bigl(\sigma({\bf r'}) C_{1}({\bf r'})+\sigma_c({\bf r'}) C_c({\bf r'})\Bigr)\label{c1}\\
C_{2}({\bf r}) & = & \int_V d{\bf r'} |\langle G({\bf r},{\bf r'})\rangle|^2 \Bigl(\sigma^*({\bf  r'}) C_{2}({\bf r'})+\Bigr.\nonumber\\
& & \ \ \ \ \ \ \ \Bigl.+\tilde\sigma({\bf r'}) C_{1}({\bf r'})+\tilde{\sigma}_c({\bf r'})C_c({\bf r'})\Bigr)\ \ \ \label{c2}
\end{eqnarray}
from which the crossed
      bistatic coefficient (i.e. the height of the coherent backscattering cone) results as:
\begin{eqnarray}
\Gamma_C   & = &  \int_V \frac{d{\bf r}}{4\pi A}C_c^*({\bf r})\times\nonumber\\
& & \times \Bigl((\sigma_c^*({\bf r})+\tilde{\sigma}_c^*({\bf r}))C_1({\bf r})+\sigma_c^*({\bf r})
C_2({\bf r})\Bigr)\label{gammac}
\end{eqnarray}
The coherent backscattering enhancement factor is then defined as $\eta=(\Gamma_C+\Gamma_L)/\Gamma_L$, i.e. the total backscattered intensity
divided by the incoherent background.
Writing the iterative solutions of Eqs.~(\ref{c1},\ref{c2}) formally as $C_1=\sum_n (|G|^2\sigma)^n|G|^2\sigma_c C_c$, and similarly for 
$C_2$,
it can be checked (using $\sigma^*-\tilde{\sigma}\in\mathbb R$) that $\Gamma_C\in\mathbb R$, as it should be.

In the expression for $C_c({\bf r})$, we have neglected events where the coherent mode is affected by a propagation event $\tau_c$. Furthermore, Eqs.~(\ref{cc}-\ref{gammac}) are only valid in exact backscattering direction ${\bf k_D}=-{\bf k_L}$. An exact version of the crossed transport equations, where these assumptions are relaxed, can be found in Appendix \ref{sctrans}.

\subsection{Comparison between  crossed and  ladder intensity}

The interpretation of Eqs.~(\ref{c1}) is as follows: we start with the crossed intensity $C_1({\bf r'})$, or the coherent intensity $C_c({\bf r'})$, at point $\bf r'$ on the right hand side. Here, it undergoes a scattering event described by the corresponding cross section $\sigma({\bf r'})$ or $\sigma_c({\bf r'})$, respectively. After being scattered, the crossed intensity propagates to ${\bf r}$ according to the average Green's function $|\langle G({\bf r},{\bf r'})\rangle|^2$, see Eq.~(\ref{green}), which finally yields $C_1({\bf r})$ on the left hand side. A similar structure applies to Eq.~(\ref{c2}).  

For comparison
      with the ladder background $\Gamma_L$, it is helpful to bring the ladder transport equation for $I_d({\bf r})$, see Eq.~(\ref{id}), into the same form as Eqs.~(\ref{c1},\ref{c2}). For this purpose, we write
\begin{equation}
K=\sigma^{(d)} I_d+\sigma^{(d)}_c I_c\label{sigmadc}
\end{equation}
with suitably defined "diffuson cross sections" $\sigma^{(d)}$ and $\sigma_c^{(d)}$, see Eq.~(\ref{sigmad}) below. Then, Eq.~(\ref{id}) turns into the desired form
\begin{eqnarray}
I_d({\bf r})  & =  & \int_V d{\bf
  r'}|\langle G({\bf r},{\bf r'})\rangle|^2\times\nonumber\\
& & \times \Bigl(\sigma^{(d)}({\bf r'}) I_d({\bf r'})+\sigma^{(d)}_c({\bf r'}) I_c({\bf r'})\Bigr)\label{id2}
\end{eqnarray}
Making use of the intensity statistics defined by
Eqs.~(\ref{pi},\ref{pi2}), we can show
that the following definition: 
\begin{equation}
\sigma^{(d)} =  \sigma-\tilde{\sigma}^*\label{sigmad}
\end{equation}
and, similarly, $\sigma_c^{(d)}=\sigma_c-\tilde{\sigma}^*_c$, fulfills the required property, Eq.~(\ref{sigmadc}).
Eq.~(\ref{sigmad}) can be rewritten as 
\begin{equation}
\sigma^{(d)}=\kappa-\tilde{\kappa}^*={\mathcal N}\left<\frac{d|f|^2}{d I}\right>
\label{sigmad2}
\end{equation}
In the conservative case, where the polarizability $g(I)$ fulfills the optical theorem, Eq.~(\ref{optth}),
the diffuson cross sections are related to the extinction path lengths through
$\sigma^{(d)}=4\pi/\ell$ and $\sigma_c^{(d)}=4\pi/\ell_c$.

Using Eq.~(\ref{sigmad}), we can now discuss the relation between the ladder and crossed intensities as follows:
First, we note that in the linear case, where $\tilde{\sigma}=0$,
diffuson and cooperon cross sections $\sigma^{(d)}$ and $\sigma$ are equal, as expected from the
reciprocity symmetry.
This implies the well known result that, in the linear case,
the coherent backscattering cone height $\Gamma_C$ equals the background
$\Gamma_L$ (minus the single scattering contribution).

Next, we turn to the case $\tilde{\sigma}\in{\mathbb R}$. Due to
$\sigma^{(d)}\in {\mathbb R}$, this implies
also $\sigma\in{\mathbb R}$.
As we can easily see from
Eqs.~(\ref{tildekappa},\ref{tau},\ref{tildesigma}), this case is realized for a
purely imaginary nonlinear polarizability ($g\in
i{\mathbb R}$), or, in other words, if the nonlinearity modifies only
the imaginary part of the refractive index. If $\tilde{\sigma}>0$,
we find $\sigma>\sigma^{(d)}$ from Eq.~(\ref{sigmad}), i.e., the
cooperon cross section is larger than the diffuson cross section,
and, consequently, $C_1>I_d$. In addition, we have $C_2>0$, and,
consequently, also $C_1+C_2>I_d$. We conclude that the nonlinearity
enhances the cone height  relative to the background if $\tilde{\sigma}>0$. In a similar way,
we can show that $\Gamma_C/\Gamma_L$  is reduced if $\tilde{\sigma}<0$.
 
In order to find out whether $\tilde{\sigma}$ is positive or negative,
we can use the following relation, which is valid in the absence of 
coherent intensity (i.e. deep inside the bulk):
\begin{equation}
\left.{\rm Re}\{\ell\tilde{\sigma}\}\right|_{I_c=0}=
\left. \frac{I_d}{2}\frac{d(\ell\sigma^{(d)})}{dI_d}\right|_{I_c=0}\label{rel2}
\end{equation}
The proof of Eq.~(\ref{rel2}) is demonstrated in Appendix~\ref{proofrel2}.
From Eq.~(\ref{rel2}), we predict (in the case
$\tilde{\sigma}\in{\mathbb R}$)
that the coherent backscattering cone is enhanced by the
nonlinearity, if the scaled cross section $\ell\sigma^{(d)}$ increases
as a function of the diffuse intensity, and vice versa.
Note that the condition $\ell\sigma^{(d)}=4\pi$ just expresses the optical theorem valid for the case of a conservative medium. Thus, $\ell\sigma^{(d)}<4\pi$
corresponds to an absorbing, and $\ell\sigma^{(d)}>4\pi$ to an amplifying medium. Thereby, we arrive at the following conclusion:
A coherent backscattering enhancement factor larger than the linear value $2$ (neglecting single scattering) will be observed in the case of amplifying media, if the amplification increases with increasing wave intensity $I_d$, or in the case of absorbing media, if the absorption decreases with increasing $I_d$ (absorption saturation).
In the opposite cases (gain saturation or increasing absorption), the coherent backscattering enhancement factor will be diminished with respect to the linear case. This latter prediction of our theory was confirmed by 
exact numerical simulations in \cite{wellens08}.

In the case of a general nonlinear polarizability $g$, which
also (or only) affects the real part of the refractive index,
the cooperon cross sections $\sigma$ and 
$\tilde{\sigma}$ are complex numbers with non-vanishing phase.
This implies the occurrence
of phase differences between the reversed crossed paths, and,
consequently, the height of the coherent backscattering cone is expected to decrease.

\section{Homogeneous Kerr nonlinearity}
\label{skerr}

The transport equations derived in Secs.~\ref{sladder} and \ref{scrossed} are valid for the nonlinear point scatterer model with completely general form of the
polarizability $g(I)$.  
In the present section,
we will apply the diagrammatic method to a different type of nonlinear disordered system, namely
linear scatterers embedded in a homogeneous Kerr medium.  In this case, the general structure of the transport equations will turn out to be the same as for the nonlinear point scatterer model - only the formulas for calculating the diffuson and cooperon cross sections are modified, as shown below.

\subsection{Model and transport equations}
\label{smodel2}

We consider the following nonlinear Helmholtz equation, describing
propagation of light in a medium with randomly fluctuating refractive index $\delta\epsilon({\bf r})$
and constant Kerr nonlinearity $\alpha$:
\begin{equation}
\Delta E({\bf r})+k^2\Bigl(1+\delta\epsilon({\bf r})+\alpha|E({\bf
  r})|^2\Bigr)E({\bf r})=0\label{hom}
\end{equation}
We note that an equivalent equation (known as Gross-Pitaevskii equation) also describes propagation of Bose-Einstein condensates in disordered potentials \cite{bec}. Concerning the disorder potential $\delta\epsilon({\bf r})$,
we assume, for simplicity, short range correlations, i.e.
$\langle\delta\epsilon({\bf r})\delta\epsilon({\bf r'})\rangle\neq  0$ only if  $k|{\bf r}-{\bf r'}|\ll 1$.
Then, $\delta\epsilon({\bf r})$ induces isotropic scattering, which is characterized by a single parameter, i.e. the mean free path $\ell_0$.

The dilute medium approximation for Eq.~(\ref{hom}) is valid if
$k\ell_0\gg 1$ and $(\alpha I)^2 k\ell_0\ll 1$. The latter condition ensures that scattering due to fluctuations of the nonlinear refractive index  can be neglected \cite{spivak03}. This 
is automatically fulfilled if we assume that we are in the 
regime  where Eq.~(\ref{hom}) has a unique stable solution.  According to
\cite{skipetrov}, this is the case (for $\alpha\in{\mathbb R}$) if
$(\alpha I)^2 b^2(k\ell_0+b)<1$, with $b$ the optical thickness. 

To apply the diagrammatic method to Eq.~(\ref{hom}), we first rewrite it as an integral equation:
\begin{eqnarray}
E({\bf r}) & = & e^{i{\bf k_L}\cdot{\bf r}}+k^2\int d{\bf r}~G_0({\bf r},{\bf r'})\times
\nonumber\\
& & \times \Bigl(\delta\epsilon({\bf r'})+\alpha |E({\bf r'})|^2\Bigr)E({\bf r'})\label{homint}
\end{eqnarray}
which solves Eq.~(\ref{hom}) for the case of an incoming plane wave. By iteration, we now obtain the nonlinear Born series in a similar way as for Eq.~(\ref{model1}). When performing the ensemble average, some complications arise concerning the nonlinear scattering events represented by the term $\alpha |E({\bf r})|^2 E({\bf r})$ in Eq.~(\ref{homint}). Since their position ${\bf r}$ is not restricted to a discrete set of dilute point scatterers, we must, in principle, take into account correlations of the form 
$|\alpha|^2|\langle  |E({\bf r_1})|^2 E({\bf r_1})|E({\bf r_2})|^2 E^*({\bf r_2})\rangle$ with ${\bf r_1}$ close to ${\bf r_2}$. It turns out, however, that we may neglect these terms if we assume $(\alpha I)^2 k\ell_0\ll 1$, as mentioned above. Since only linear scattering events remain, we immediately get the following results valid for linear point scatterers:
\begin{eqnarray}
\sigma^{(d)} = \sigma_c^{(d)} = \kappa  =\kappa_c & = & \frac{4\pi}{\ell_0}\label{sigmahom}\\
\tilde{\kappa} = \tilde{\kappa}_c & = & 0\label{tildekappahom}
\end{eqnarray} 

Concerning nonlinear average propagation, we recover the same equations (\ref{index},\ref{nc1}) for the refractive indices and Eq.~(\ref{tau}) for the crossed propagation diagram as for the nonlinear point scatterer, if we replace, in these equations, the density of scatterers times the polarizability by the corresponding expression for the Kerr nonlinearity,
i.e. ${\mathcal N}g(I)\to k^2\alpha I$. Adding
the linear index $n_0-1=i/(2k\ell_0)$, we obtain in total:
\begin{eqnarray}
n-1 & = & \frac{i}{2k\ell_0}+\alpha (I_c+I_d)\label{nhom}\\
n_c-1 & = & \frac{i}{2k\ell_0}+\alpha \left(\frac{I_c}{2}+I_d\right)\label{nchom}
\end{eqnarray}
with corresponding extinction path lengths $\ell=1/(2k{\rm Im}n)$ and $\ell_c=1/(2k{\rm Im}n_c)$, and
\begin{equation}
\tau=\tau_c=-ik\alpha^*\label{tauhom}
\end{equation}
The above expressions, Eqs.~(\ref{sigmahom}-\ref{tauhom}), can now be inserted into the transport equations for the ladder and crossed intensity, Eqs.~(\ref{id},\ref{ic}) and Eqs.~(\ref{ccex}-\ref{coex}).
Also the approximate form, Eqs.~(\ref{cc}-\ref{c2}), as well as the relation (\ref{rel2}) between diffuson and cooperon cross section remain equally valid for the homogeneous nonlinear medium, with the cooperon cross sections $\sigma$ and $\tilde{\sigma}$ given by
Eqs.~(\ref{sigma},\ref{tildesigma}). 

\subsection{Results for conservative nonlinearity ($\alpha\in{\mathbb R}$)}

First, we examine the case $\alpha\in{\mathbb R}$, where energy is conserved (more precisely:
the intensity flux ${\bf j}={\rm Re}\{iE \nabla E^*\}$ fulfills the conservation law $\nabla {\bf j}=0$). In this case, we see from Eq.~(\ref{sigmahom}) and Eqs.~(\ref{nhom},\ref{nchom}) that the quantities $K$ (or, equivalently $\sigma^{(d)}$ and $\sigma^{(d)}_c$), $\ell$ and $\ell_c$, which enter into the ladder transport equations (\ref{id},\ref{ic}), are unaffected by the nonlinearity $\alpha$. Hence, the average diffuse and coherent intensities $I_d({\bf r})$ and $I_c({\bf r})$, as well as the background component $\Gamma_L$ of the scattered intensity, see Eq.~(\ref{gammal}), are the same as in the linear case. This is not surprising, since for $\alpha\in{\mathbb R}$, the nonlinearity modifies only the real part of the refractive index, which, in the case of a dilute medium, has a negligible influence on incoherent wave transport.

However, the nonlinearity does affect interference properties like coherent backscattering, which are described by crossed diagrams. To see this, we first deduce from Eqs.~(\ref{sigmahom}-\ref{tauhom})  the cooperon cross section for $\alpha\in{\mathbb R}$, and obtain:
\begin{eqnarray}
\sigma({\bf r}) =  \sigma_c({\bf r}) & = & \frac{4\pi}{\ell_0}\left[1-ik\ell_0\alpha \left(I_c({\bf
    r})+I_d({\bf r})\right)\right],\label{sigmahomcons}\\
\tilde{\sigma}({\bf r}) =\tilde{\sigma}_c({\bf r}) & = & 4\pi ik\alpha \left(I_c({\bf r})+I_d({\bf r})\right)
\end{eqnarray}
Of special interest is the fact that ${\rm Re}\tilde{\sigma}=0$. From Eq.~(\ref{rel2}), we see that this is generally true 
(at least inside the bulk where $I_c=0$)
for conservative media, where $\ell\sigma^{(d)}=4\pi$. Using the condition ${\rm Re}\tilde{\sigma}=0$, it can be shown (again with help of the iterative solutions for $C_1$ and $C_2$) that:
\begin{equation}
\Gamma_C={\rm Re}\left\{\int_V \frac{d{\bf r}}{4\pi A}C_c^*({\bf r})\sigma({\bf r}) C_1({\bf r})\right\}\label{gammachom}
\end{equation}
If $C_c\in\mathbb R$ (i.e. if the difference between $n({\bf r})$ and $n_c({\bf r})$ is neglected), Eq.~(\ref{gammachom}) can also be shown in a simpler way,
since then ${\rm Re}C_2=0$ and ${\rm Im}C_2=-{\rm Im}C_1$ follow directly by
splitting Eqs.~(\ref{c1},\ref{c2}) into real and imaginary parts. 

Due to Eq.~(\ref{gammachom}), solving the transport equation for $C_1$, Eq.~(\ref{c1}), is sufficient for calculating the coherent backscattering enhancement. As already mentioned, this equation is very similar to the ladder transport equation, Eq.~(\ref{id2}), with the difference that the cross section, Eq.~(\ref{sigmahom}),  is a complex quantity $\sigma=|\sigma|\exp(i\phi)$. This can be interpreted as follows: at each scattering event, the nonlinearity introduces a certain phase difference $\phi$ between the two reversed paths, which depends on the average intensity
$\langle I({\bf r})\rangle=I_d({\bf r})+I_c({\bf r})$ multiplied by $k\ell_0\alpha$. As a result of this dephasing mechanism, the height of the coherent backscattering cone decreases with respect to the linear case. Note that the sign of $\alpha$ plays no role, i.e. we obtain the same value of $\Gamma_C$  for focussing ($\alpha>0$) and defocussing ($\alpha<0$) nonlinearity, respectively. This also results in an initial decrease quadratic in 
$\alpha$. 

In the conservative case, we hence recover the familiar picture of coherent backscattering as an interference between only two reversed amplitudes - however with a phase difference induced by the nonlinearity. At first sight, it might be surprising that the nonlinearity does not affect both counterpropagating amplitudes in the same way. This can be understood only by taking into account the many-wave interference character of nonlinear coherent backscattering. In the conservative case, every nonlinear event gives rise to a purely imaginary factor, i.e.
$\tau=-ik\alpha$ or $\tau^*=+ik\alpha$, such that most of the many interference terms cancel each other. At the end - as a consequence of the forbidden diagrams, see Sec.~\ref{sforbidden} - it turns out that only a single interference term remains, where the nonlinearity 
modifies only one of the two counterpropagating amplitudes $\psi$ or $\psi^*$, but not the other one.

In the limit $\alpha k\ell_0 I\ll 1$, we find from Eq.~(\ref{sigmahomcons}) the phase difference $\Delta\phi=M k\ell_0\alpha I$ for a path undergoing $M$ scattering events. In the case of a slab geometry
(finite length $L=b\ell_0$ in $z$ direction, infinite extension in $x$ and $y$ direction)
the mean number $\langle M\rangle$ of  scattering events for diffuse reflection is proportional to the optical thickness $b$ of the scattering medium (in contrast to diffuse transmission, where $\langle M\rangle\propto b^2$). From this, we predict a significant reduction of the
coherent backscattering peak if 
$b k\ell_0\alpha I\simeq 1$. Note that, if $b$ and $k\ell_0$ are large, already a very tiny nonlinearity is sufficient to observe a large impact on the CBS peak, which, in particular, still lies in  the regime
$(\alpha I)^2 b^2(k\ell_0+b)<1$, where, according to \cite{skipetrov}, the wave equation (\ref{hom}) has a unique stable solution.

\begin{figure}
\centerline{\includegraphics[angle=0,width=8cm]{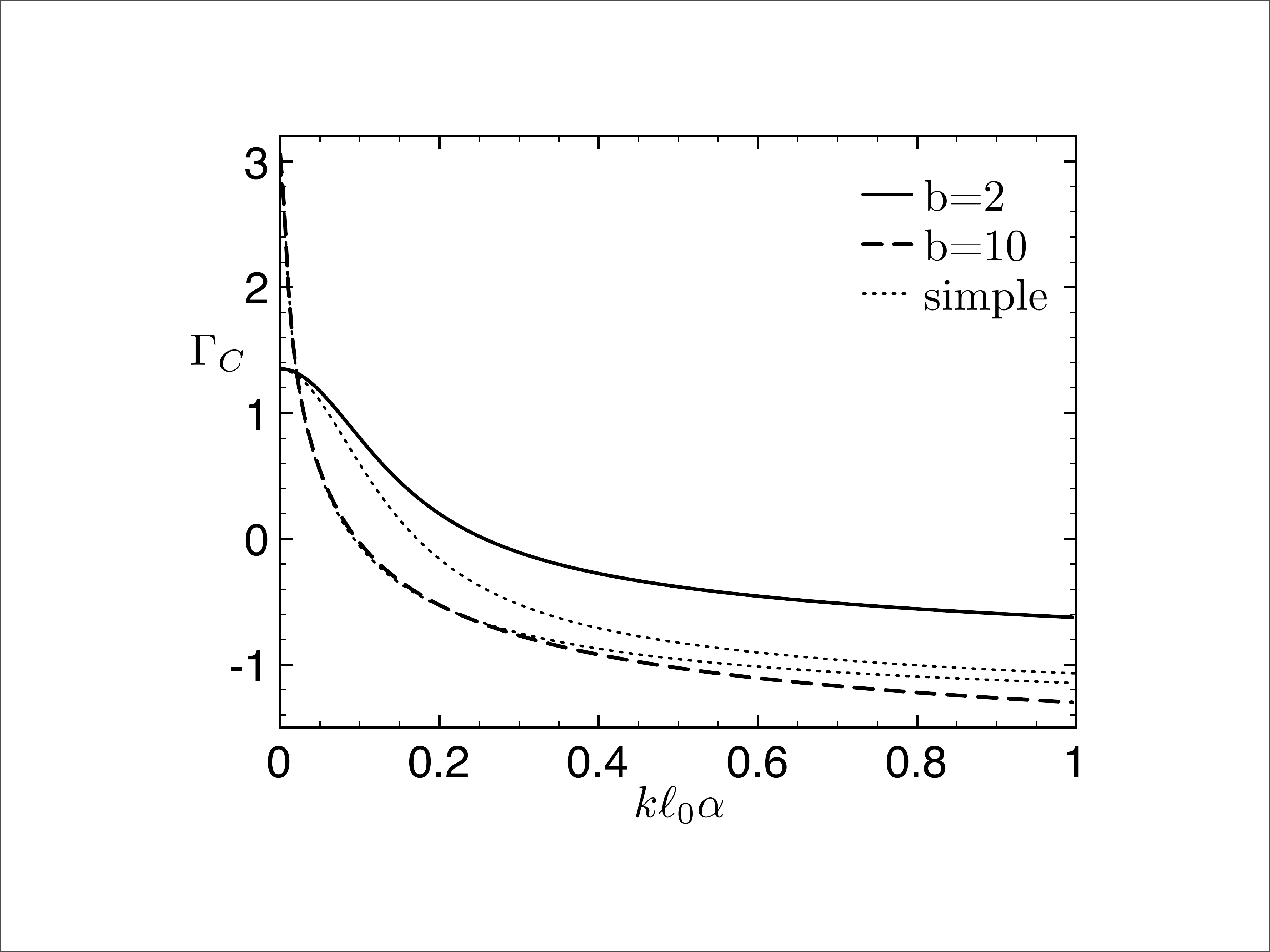}}
\caption{Coherent backscattering cone height $\Gamma_C$ as a function of the nonlinearity strength
$k\ell_0\alpha$ for a slab with optical thickness $b=2$ (solid line) and $b=10$ (dashed line).
The initial decrease of the cone height due to dephasing between reversed paths
is faster for larger optical thickness. At large nonlinearities,
the cone height $\Gamma_C$ assumes negative values, indicating destructive interference.
The thick lines result from the complete summation of crossed diagrams, see
Appendix~\ref{sctrans}, whereas the thin dotted lines correspond to the simplified version,
Eqs.~(\ref{cc}-\ref{gammac}). \label{figcons}} 
\end{figure}

The above conclusions are confirmed by Fig.~\ref{figcons}, showing the results obtained from the exact transport equations, Eqs.~(\ref{ccex}-\ref{gammacex}), compared to their approximate form, Eqs.~(\ref{cc}-\ref{gammac}), for a slab geometry with optical thickness $b=2$ and $b=10$, respectively.
As expected, the approximate form agrees very well 
with the exact result  for large optical thickness and small nonlinearity. Furthermore, the initial decrease, quadratic in $\alpha$,
of the cone height occurs much faster for $b=10$ than for $b=2$, as predicted above.
Remarkably, the cone height assumes, in both cases, negative values for large nonlinearities, corresponding to destructive interference in backscattering direction. This property has recently been observed also in numerical studies of the 2D Gross-Pitaevskii
equation \cite{hartung08}.

\subsection{Results for absorbing/amplifying nonlinearity ($\alpha\in i{\mathbb R}$)}

Next, we discuss the case $\beta=ik\ell_0\alpha$, with $\beta\in{\mathbb R}$. Then, $\beta>0$ corresponds to an amplifying, and $\beta<0$ to an absorbing medium. 
According to Eqs.~(\ref{nhom},\ref{nchom}), we find for the extinction path lengths:
\begin{eqnarray}
\frac{1}{\ell} & = & \frac{1}{\ell_0}\left(1-2\beta (I_c+I_d)\right)\\
\frac{1}{\ell_c} & = & \frac{1}{\ell_0}\left(1-\beta (I_c+2I_d)\right)
\end{eqnarray}
Note that, for $\beta>0$ sufficiently large, the extinction path length may, in principle, become negative (as a consequence of strong amplification). However, as we will find below, stationary solutions of the 
nonlinear transport equations typically exist only for small $\beta$. Therefore, we assume $\beta\ll 1$, and hence
$\ell>0$ in the following.

\begin{figure}
\centerline{\includegraphics[angle=0,width=8cm]{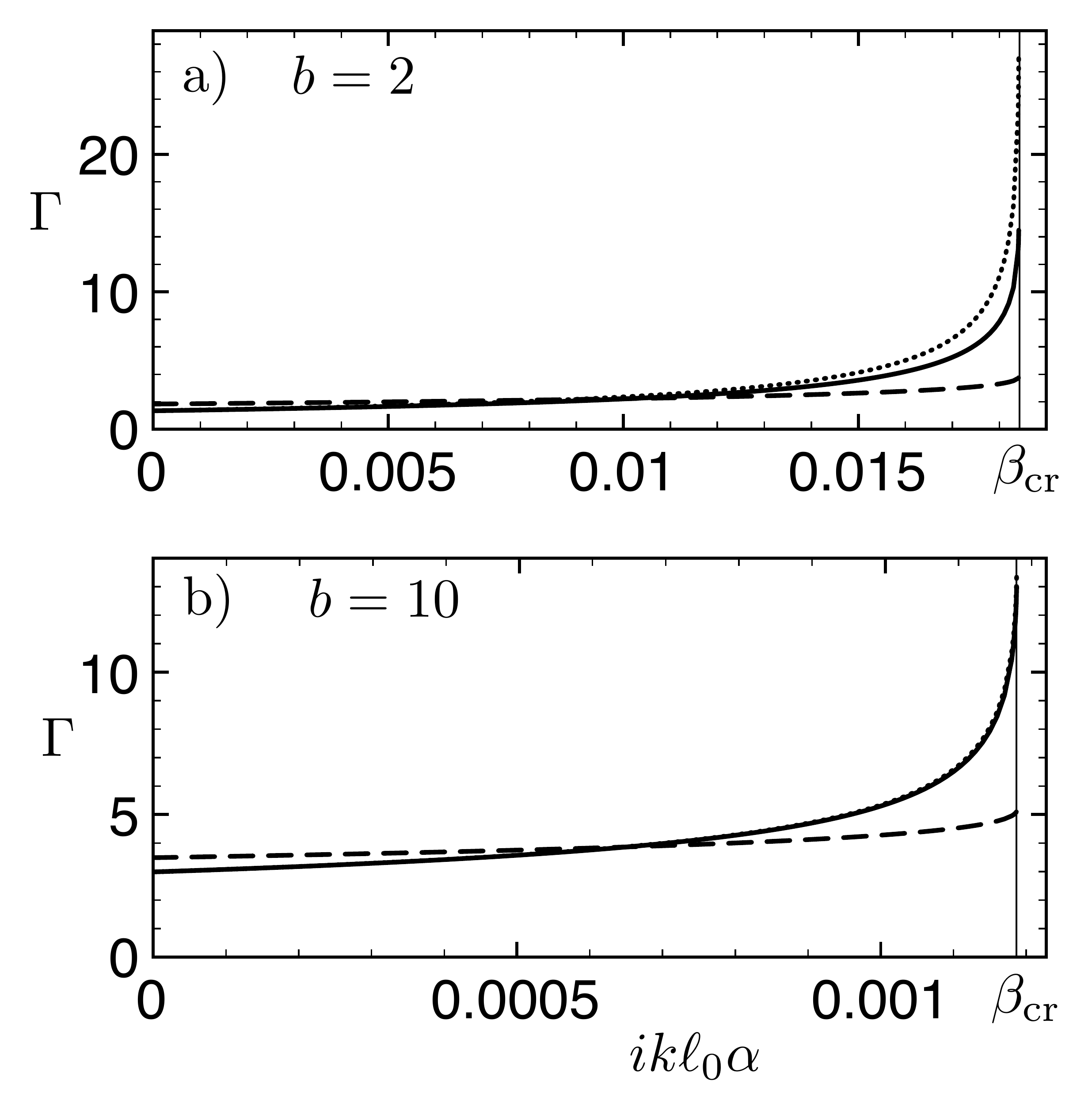}}
\caption{Background $\Gamma_L$ (dashed lines) and cone height $\Gamma_C$ (exact version: solid lines, simplified version: dotted lines)
as a function of $\beta=ik\ell_0\alpha$ (amplifying nonlinearity), for a slab with optical thickness a) $b=2$
and b) $b=10$. The cone height is more strongly affected  by nonlinear amplification than the background.
Close to the critical value $\beta_{\rm cr}$ of the nonlinearity, the cone height 
exceeds the background intensity by more than a factor 3.
\label{figimag}} 
\end{figure}

By inserting Eqs. (\ref{sigmahom},\ref{tildekappahom},\ref{tauhom}) into the definitions of $\sigma$ and $\tilde{\sigma}$, Eqs. (\ref{sigma},\ref{tildesigma}), we obtain:
\begin{eqnarray}
\sigma=\sigma_c & = & \frac{4\pi}{\ell_0} \frac{1-\beta (I_c+I_d)}{1-2\beta (I_c+I_d)}\\
\tilde{\sigma}=\tilde{\sigma}_c & = & \frac{4\pi}{\ell_0} \frac{\beta (I_c+I_d)}{1-2\beta (I_c+I_d)}
\end{eqnarray}
From the sign of $\tilde{\sigma}$, we can now conclude, following
the discussion in Sec.~\ref{crossedtrans}, that
the CBS cone height increases for $\beta>0$, and decreases for $\beta<0$. This is confirmed by
Fig.~\ref{figimag}, where the background intensity $\Gamma_L$ and cone height $\Gamma_C$ are plotted as a function of $\beta$, for a slab with optical thickness $b=2$ and $b=10$, respectively.
Again, the exact solutions for $\Gamma_C$ (solid lines) are compared with the simplified version
(dotted lines), which fits well in the case of large optical thickness and small nonlinearity, as expected.
More importantly, however,
we see that the cone height is more strongly amplified by the nonlinearity than the background,
such that $\Gamma_C$ may exceed $\Gamma_L$ by more than a factor three. This is a direct
evidence for the multi-wave interference character of CBS in the nonlinear regime, as mentioned
in Sec.~\ref{sscrossed}.

However, we note that the backscattering cone height $\Gamma_C$ does not become arbitrarily large.
When increasing the amplifying nonlinearity beyond a certain critical value  $\beta_{\rm cr}$,
we find that, for $\beta>\beta_{\rm cr}$, the stationary nonlinear ladder transport equations (\ref{id},\ref{ic}) have no solution.
This behavior can be understood if we approximate Eq.~(\ref{id}) by a nonlinear diffusion equation. For this
purpose, we assume that the average intensity $I(z)=I_c(z)+I_d(z)$ varies slowly on the
length scale of a mean free path, i.e. $\ell_0 I'(z)\ll 1$ and $\ell_0^2I''(z)\ll 1$. Additionally, we assume
$\beta\ll 1$. Expanding Eq.~(\ref{id}) up to lowest non-vanishing order in those small parameters,
we obtain:
\begin{equation}
\frac{\ell_0^2}{3}I''+2\beta I^2+e^{-z/\ell_0}=0\label{diffeq}
\end{equation}
where we have approximated $I_c(z)\simeq \exp(-z/\ell_0)$. The boundary conditions read:
$I(0)-z_0I'(0)=0$ and $I(L)+z_0I"(L)=0$, with $L=b\ell_0$ the thickness of the slab,
and the extrapolation length $z_0\simeq 0.71\ell_0$ \cite{chantry87}. From the solution of Eq.~(\ref{diffeq}),
the background intensity is obtained as $\Gamma_L=\int_0^L dz I(z) \exp(-z/\ell_0)$.

\begin{figure}
\centerline{\includegraphics[angle=0,width=8cm]{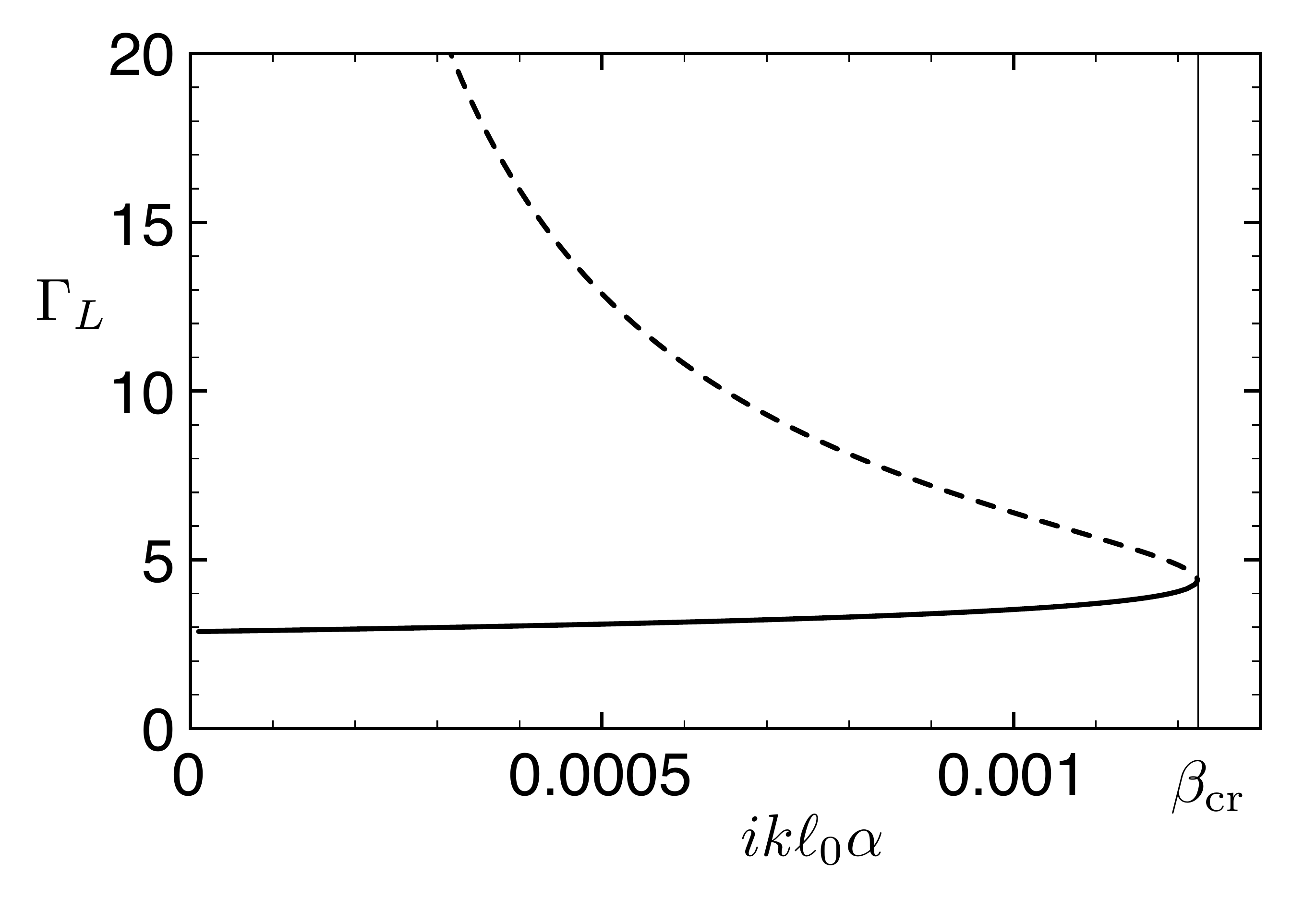}}
\caption{Background intensity $\Gamma_L$ as a function of $\beta=ik\ell_0\alpha$ (amplifying nonlinearity), obtained from the diffusion equation (\ref{diffeq}) for a slab with optical thickness $b=10$. For $0<\beta<\beta_{\rm cr}$, two solutions are found. The lower branch (solid line) reproduces the iterative solution of the ladder transport equations (\ref{id},\ref{ic}), cf. the dashed line in Fig.~\ref{figimag}(b), whereas the upper branch (dashed line) corresponds to a physically unstable solution. No positive solution of Eq.~(\ref{diffeq}) exists for  $\beta>\beta_{\rm cr}$.
\label{figdiffgl}} 
\end{figure}

For small $\beta>0$, the nonlinear diffusion equation (\ref{diffeq}) has two different solutions, see Fig.~\ref{figdiffgl}. One of them (the solid line) reproduces approximately
the iterative solution of the ladder transport equations shown in Fig.~\ref{figimag}.
On the other hand, the iteration of the ladder transport equations diverges if the initial condition is chosen close to the other solution (dashed line in Fig.~\ref{figdiffgl}). We suspect that the lower branch (solid line) corresponds to a dynamically stable and the upper branch (dashed line) to a dynamically unstable solution of the nonlinear diffusion equation.

Finally, for $\beta>\beta_{\rm cr}$, the nonlinear term $2\beta I^2$ in the diffusion equation (\ref{diffeq}) enforces a too large negative curvature, $I''(z)<0$, such that no solution exists which is positive,
$I(z)\geq 0$, in the entire interval $0\leq z\leq L$, and which satisfies the boundary conditions mentioned above. Hence, in this regime, our assumption that the wave equation (\ref{hom}) possesses a stable stationary solution (see Sec.~\ref{ssstable}) cannot be fulfilled. Instead, an explicitly time-dependent treatment of the scattering problem would be required.

\section{Conclusion}
\label{concl}

We have presented a diagrammatic theory of nonlinear transport and coherent backscattering for dilute, disordered media with weak nonlinearity.
The theory was applied to two different models: nonlinear point scatterers and linear scatterers embedded in a homogeneous nonlinear medium.
The relevant nonlinear effects captured by the theory are the following: first, scattering cross section, refractive index, and mean free path depend 
on the average wave intensity and its statistical fluctuations. Furthermore, the refractive indices are different for coherent and diffuse waves, respectively. These properties are relevant for incoherent transport of the average wave intensity. 
Concerning properties of coherent transport, the most important conclusion is that, due to nonlinear coupling between different partial waves, coherent backscattering turns into
a multi-wave interference effect - instead of two-wave interference, as in the linear regime.

How the multi-wave interference character affects coherent backscattering quantitatively, depends, of course, on the phases between the various interfering amplitudes. Complete constructive interference - and hence nonlinear enhancement of the coherent backscattering effect 
 -  is found for the case of amplifying nonlinearity. For a conservative nonlinearity,
we find effectively only two interfering amplitudes, however with a phase difference proportional to the nonlinearity strength times the average intensity integrated along the whole scattering path. In this case, the coherent backscattering is suppressed (or even inverted) by the nonlinearity. 

The nonlinear enhancement of coherent backscattering predicted in the case of amplifying nonlinearity opens interesting perspectives
concerning the phenomenon of Anderson localization, which (in the linear case) is expected to occur in the regime $k\ell\leqsi 1$ of strongly
scattering media. Is Anderson localization affected in a similar way as coherent backscattering, i.e., can it be suppressed or enhanced, 
depending on the type of nonlinearity? This is an interesting problem to be investigated in future research. 
 
 \acknowledgements
We thank Dominique Delande, Michael Hartung, Christian Miniatura, Cord M\"uller and Peter Schlagheck  for fruitful
discussions. T.W. acknowledges support from the DFG.

\appendix

\section{Exact summation of crossed diagrams}
\label{sctrans}

In the approximate version of the crossed transport equations, Eqs.~(\ref{cc}-\ref{c2}), the crossed scattering and propagation diagrams, Figs.~\ref{crossedbb}(a-d) and (e,f), are treated on an equal footing, such that their combined effect is simply expressed by the cooperon cross sections $\sigma$ and $\tilde{\sigma}$,
Eqs.~(\ref{sigma},\ref{tildesigma}). In this appendix, we demonstrate the exact calculation of the nonlinear propagation diagrams, and examine the conditions under which the approximate version is expected to be valid.

\subsection{Crossed propagation diagram}

\begin{figure}
\centerline{\includegraphics[width=8.5cm]{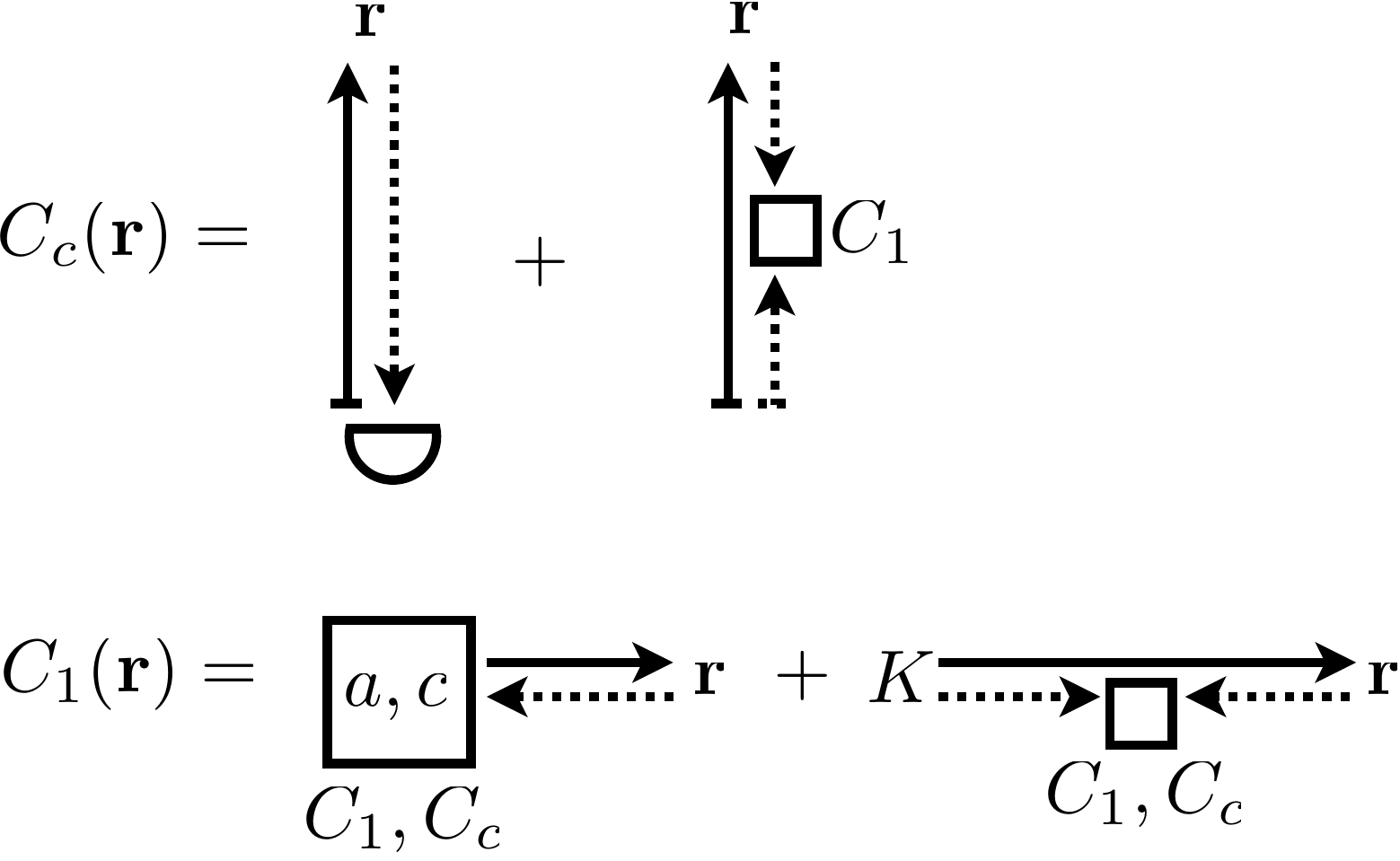}}
\caption{Graphical representation of the transport equations for $C_1$ and $C_c$, Eqs.~(\ref{ccex},\ref{c1ex}). The letters \lq a\rq\ and \lq c\rq\ refer to the crossed scattering building block from Fig.~\ref{crossedbb}, i.e. they represent the quantity $\kappa_a+\kappa_c=\kappa$, see Eq.~(\ref{kappa}). 
 \label{ctrans1}} 
\end{figure}

Let us first look at diagram Fig.~\ref{crossedbb}(e), which corresponds to the quantity $\tau$, Eq.~(\ref{tau}).  Similarly as in the ladder case, Fig.~\ref{ladderbb}(b), we denote the positions corresponding to the bottom and the top part of the diagram, which are connected by the vertical solid arrow, as ${\bf r_1}$ and ${\bf r_2}$, respectively. The position of the scattering event $\Box$ is denoted as ${\bf r_3}$.
Due to the condition of stationary phase, see Eq.~(\ref{1storder}), ${\bf r_3}$ must be situated on the straight line between ${\bf r_1}$ and ${\bf r_2}$, i.e. ${\bf r_3}=s{\bf r_1}+(1-s){\bf r_2}$,
with $0\leq s\leq 1$.
The pair of reversed arrows on the right hand side represents an incoming crossed intensity, which we denote by
$C_{\rm in}({\bf r_3})$. From ${\bf r_1}$ (the bottom), a ladder intensity is emitted, which is described by
$K({\bf r_1})$. In total, the diagram Fig.~\ref{crossedbb}(e) represents the following expression for the
crossed intensity at ${\bf r_2}$:
\begin{eqnarray}
C^{(e)}({\bf r_2}) & = & \int_Vd{\bf r_1}K({\bf r_1})|\left<G_{12}\right>|^2\times\nonumber\\
& & \times  r_{12}\int_0^1 ds \tau({\bf r_3}) C_{\rm in}({\bf r_3})=\label{ce1}\\
& = & \int_V d{\bf r_3} \tau({\bf r_3})C_{\rm in}({\bf r_3})|\left<G_{23}\right>|^2\times\nonumber\\
& & \times \int_0^{\rho_1^{(\rm max)}}d\rho_1 e^{-\rho_1/\ell} K({\bf r_1})\label{ce2}
\end{eqnarray}
where ${\bf r_1}={\bf r_3}+\rho_1 ({\bf r_3}-{\bf r_2})/r_{32}$, and the upper limit $\rho_1^{(\rm max)}$ is determined by the condition that $\bf r_1$ must lie inside the sample volume $V$. Eq.~(\ref{ce2}) results from Eq.~(\ref{ce1}) by a change of variables $({\bf r_1},s)\to({\bf r_3},\rho_1)$. In the last line, we use a short-hand notation
$\exp(-\rho_1/\ell):=\exp\left(-\rho_1\int_0^1dt/\ell({\bf r_3}-t{\bf r_3}+t{\bf r_1})\right)$
for the average of $1/\ell({\bf r})$ over the line of length $\rho_1$ connecting ${\bf r_1}$ and ${\bf r_3}$.

\subsection{Equations for $C_1$}

According to our discussion in Sec.~\ref{sforbidden}, we recall that diagram Fig.~\ref{crossedbb}(e) contributes to the crossed intensity of type $C_1$. The remaining diagrams contributing to $C_1$ are
Fig.~\ref{crossedbb}(a) and (c), which are described by the quantity $\kappa$, see Eq.~(\ref{kappa}), as follows:
\begin{equation}
C^{(a+c)}({\bf r_2})=\int_V d{\bf r_3}|\left<G_{23}\right>|^2\kappa({\bf r_3}) C_{\rm in}({\bf r_3})\label{cac}
\end{equation}
In both cases, Eq. (\ref{ce2}) and (\ref{cac}), the incoming intensity $C_{\rm in}$ may be of type
$C_1$ or $C_c$ (coherent mode). (In the latter case, $\tau$ and $\kappa$ must be replaced by
$\tau_c$ and $\kappa_c$, respectively.) Therefore, the transport equation for $C_1$ is obtained from Eqs.~(\ref{ce2},\ref{cac}) by setting $C_1=C^{(e)}+C^{(a+c)}$ and inserting $C_{\rm in}=C_1+C_c$. Similar considerations can be repeated for the calculation of the coherent mode $C_c$.
The resulting transport equations, graphically represented in Fig.~\ref{ctrans1}, read as follows:
\begin{widetext}
\begin{eqnarray}
C_c({\bf r}) & = & e^{ikn_cz}\left(e^{-ikn^*z_{\bf q}+i{\bf
      q}\cdot{\bf r}}+e^{-ikn^*z}\int_0^z dz' e^{ik(n^*-n_c^*)z'}\tau_c({\bf
    r}_{z'}) C_1({\bf r}_{z'})\right)\label{ccex}\\
C_{1}({\bf r}) & = & \int_V d{\bf r'}
\left|\langle G({\bf r},{\bf r'})\rangle\right|^2
\Biggl[\kappa({\bf r'})C_{1}({\bf
  r'})+\kappa_c({\bf r'})C_c({\bf r'})+\Biggr.\nonumber\\
  & & +\Biggl.\Bigl(\tau({\bf r'})
C_1({\bf r'})+\tau_c({\bf r'})C_c({\bf r'})\Bigr)
\int_0^{\rho_1^{(\rm max)}} d\rho_1 e^{-\rho_1/\ell}
K({\bf r_1})\Biggr]\label{c1ex}
\end{eqnarray}
\end{widetext}
with ${\bf q}={\bf k_L}+{\bf k_D}$ and ${\bf r_1}={\bf r'}+\rho_1\frac{{\bf r'}-{\bf r}}{|{\bf r'}-{\bf
      r}|}$. Furthermore, ${\bf r}_{z'}$ and $z_{\rm q}$ are defined as in Eqs.~(\ref{ec}) and (\ref{gammal}).
       Note that, in our notation for the incoming or
      outgoing waves, we have suppressed the position dependence of
      the refractive index. The latter must be averaged over the
      corresponding path, e.g.
$\exp(ikn_cz):=\exp[ik\int_0^z dz' n_c({\bf r}_{z'})]$.

\subsection{Equations for $C_2$}

\begin{figure}
\centerline{\includegraphics[width=8.5cm]{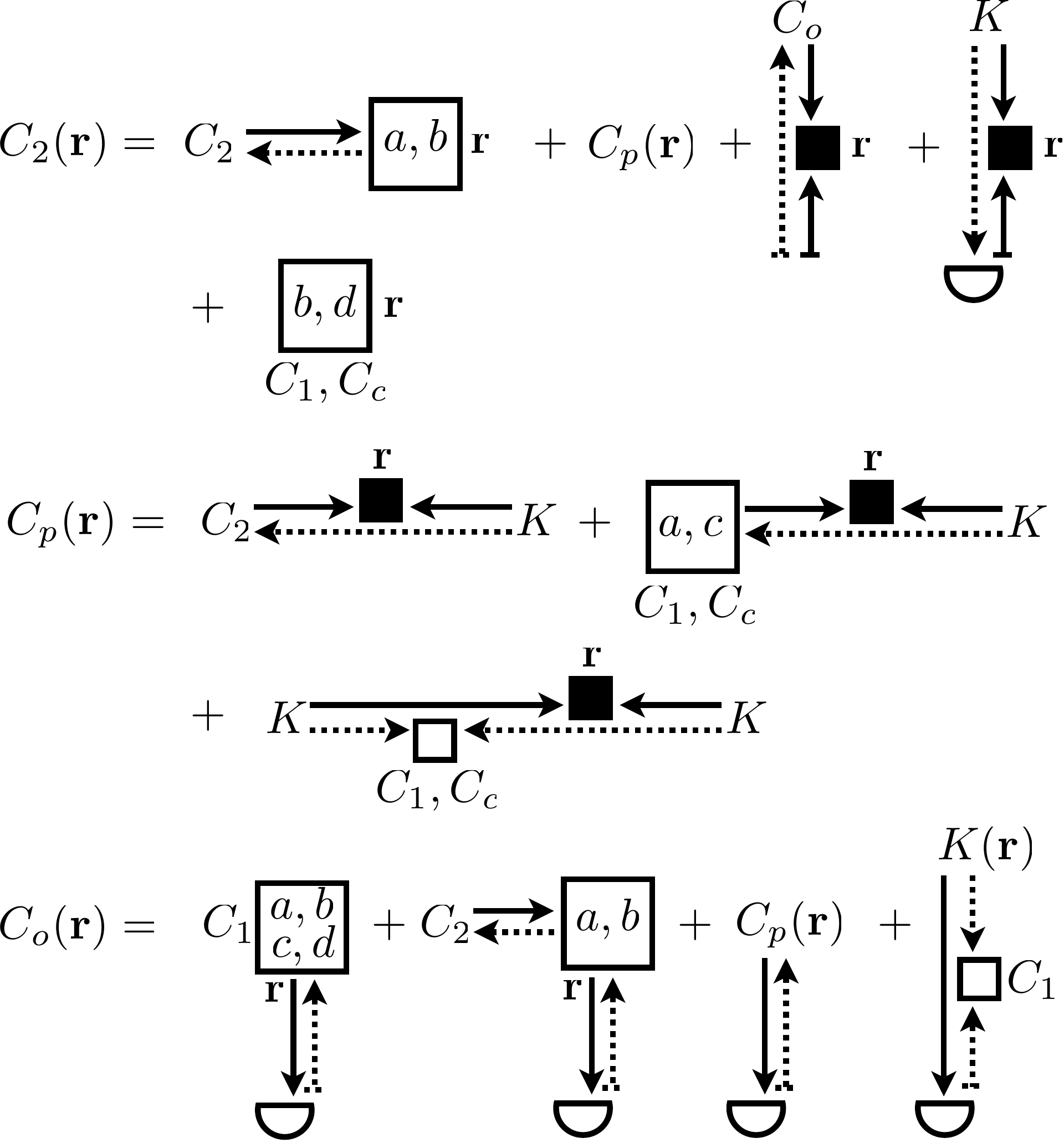}}
\caption{Graphical representation of the transport equations for the
  crossed intensity of type $C_2$, Eqs.~(\ref{c2ex}-\ref{coex}). 
\label{ctrans2}} 
\end{figure}

Concerning the crossed intensity of type $C_2$, the equation turns out to be more complicated. For example, it may happen that two nonlinear propagation events, Fig.~\ref{crossedbb}(e) and (f), occur between two scattering events, and, furthermore, all the cases must be distinguished where the incoming solid or dashed arrows do or do not originate from the coherent mode. 
Graphically, the result is shown in Fig.~\ref{ctrans2}.

In order to shorten the notation, we split the equation into three parts.
Here, $C_o$ (``o'' for
``outgoing'') encompasses all
diagrams, where the dashed arrow originates from the incoming laser and, correspondingly,
the solid arrow points towards the detector. This
quantity finally determines the coherent backscattering cone $\Gamma_C$, see
Eq.~(\ref{gammacex}).
The quantity $C_p$ 
 (``p'' for ``propagating'') contains the part of $C_2$, which originates
 from  the nonlinear propagation event Fig.~\ref{crossedbb}(f) occurring between two scattering events.

The mathematical expression of Fig.~\ref{ctrans2} yields the corresponding transport equations as follows:
\begin{widetext}
\begin{eqnarray}
C_{2}({\bf r}) & = &
\tilde{\kappa}({\bf r})C_1({\bf
  r})+\tilde{\kappa}_c({\bf r})C_c({\bf r})+\tau^*({\bf r})C_p({\bf r})+
\kappa^*({\bf r})\int_V d{\bf r'}
\left|\left<G({\bf r},{\bf r'})\right>\right|^2
 C_{2}({\bf r'})+
\nonumber\\
& & +\tau_c^*({\bf r}) e^{ik(n_c-n)z}\int_z^L dz'\Bigl(C_o({\bf
  r}_{z'})+e^{iknz'-ikn^*z_{\bf q}'+i{\bf q}\cdot{\bf r}_{z'}}K({\bf r}_{z'})\Bigr)\label{c2ex}\\
C_p({\bf r}) & = & \int_V d{\bf
  r'}\left|\left<G({\bf r},{\bf r'})\right>\right|^2
\Biggl[C_2({\bf r'})+\kappa({\bf r'})C_{1}({\bf
  r'})+\kappa_c({\bf r'})C_c({\bf
      r'}) +\Biggr.
\nonumber\\
& & \Biggl.+
  \Bigl(\tau({\bf r'})C_1({\bf r'})+\tau_c({\bf r'})C_c({\bf
    r'})\Bigr)
\int_0^{\rho_1^{(\rm max)}}
  d\rho_1 e^{-\rho_1/\ell}K({\bf
    r_1})\Biggr]
\int_0^{\rho_2^{(\rm max)}} d\rho_2 e^{-\rho_2/\ell}K({\bf r_2})\label{cpex}\\
C_o({\bf r}) & = & e^{ik(n-n_c^*)z}
  \left(\left(\kappa^*_c({\bf r})+\tilde{\kappa}^*_c({\bf r})\right)C_1({\bf
  r})+\tau^*_c({\bf r}) C_p({\bf r})+\kappa^*_c({\bf r})\int_V d{\bf r'}
\left|\left<G({\bf r},{\bf r'})\right>\right|^2
 C_{2}({\bf
    r'})\right)+\nonumber\\
& & + e^{-z/\ell} K({\bf r})\int_0^z dz' e^{ik(n^*-n_c^*)z'}
\tau_c({\bf r}_{z'})C_1({\bf r}_{z'})\label{coex}
\end{eqnarray}
\end{widetext}
with ${\bf r_1}={\bf r'}+\rho_1\frac{{\bf r'}-{\bf r}}{|{\bf r'}-{\bf
      r}|}$ and ${\bf r_2}={\bf r}+\rho_2\frac{{\bf r}-{\bf r'}}{|{\bf r}-{\bf
      r'}|}$. 
      
Finally, the height of the coherent backscattering cone reads:
\begin{equation}
\Gamma_C({\bf k_D})  =  \int_V \frac{d{\bf r}}{4\pi A} e^{ikn(z_{\bf q}-z)-i{\bf q}\cdot{\bf r}}C_o({\bf r})
\label{gammacex}
\end{equation}

\subsection{Simplified version}

It remains to be shown how the simplified version, see Eqs.~(\ref{cc}-\ref{gammac}), is derived from Eqs.~(\ref{ccex}-\ref{gammacex}). For this purpose, we assume that, when varying the position inside the sample volume on a length scale comparable to the extinction path length, the average intensity
remains approximately constant. This is the case for a medium with large optical thickness, except for the regions close to the boundary. Using this assumption, and setting $\rho_1^{\rm (max)}=\infty$ (which is again valid far away from the boundary), we simplify the integral over $\rho_1$ in Eq.~(\ref{c1ex}) as follows: 
\begin{equation}
\int d\rho_1 \exp(-\rho_1/\ell)K({\bf r_1})\simeq K({\bf r'})\ell({\bf r'})\label{intsimpl}
\end{equation}
This yields the simplified equation for $C_1$, see Eq.~(\ref{c1}), with the cooperon cross section $\sigma$, Eq.~(\ref{sigma}).
In the equation for the coherent mode $C_c$, Eq.~(\ref{ccex}), we neglect the term containing $\tau_c$, 
and obtain Eq.~(\ref{cc}). This is justified for a relatively weak nonlinearity, where a nonlinear event is unlikely to occur on the length scale $\ell$ defined by linear scattering. 

Using the same argument, we neglect the term containing $\tau_c^*$ at the end of Eq.~(\ref{c2ex}). Thereby, the $C_o$-part drops out of the equation for $C_2$, whereas the $C_p$-part is approximated by
\begin{eqnarray}
C_p({\bf r}) & \simeq  & \int_V d{\bf r'} \left|\left<G({\bf r},{\bf r'})\right>\right|^2 \Bigl(C_2({\bf r'})+\sigma({\bf r'}) C_1({\bf r'})\Bigr.\nonumber\\ & & \ \ \ \ \ \ \Bigl.+\sigma_c({\bf r'}) C_c({\bf r'})\Bigr)K({\bf r})\ell({\bf r})\simeq\\
& \simeq & K({\bf r})\ell({\bf r})[C_1({\bf r})+\tilde{C}_2({\bf r})]
\end{eqnarray}
where Eq.~(\ref{c1}) was used, and we defined
\begin{equation}
\tilde{C}_2({\bf r})=\int_V d{\bf r'} \left|\left<G({\bf r},{\bf r'})\right>\right|^2 C_2({\bf r'})
\end{equation}
Now, transforming Eq.~(\ref{c2ex}) into an equation for $\tilde{C}_2$ yields Eq.~(\ref{c2}).
Finally, the bistatic coefficient, Eq.~(\ref{gammac}) is obtained from Eqs.~(\ref{coex},\ref{gammacex}) under neglect of the last term (the one containing $\tau_c$) in Eq.~(\ref{coex}).

In summary, the simplified transport equations, Eqs.~(\ref{cc}-\ref{gammac}), are valid in the case of large optical thickness ($b\gg 1$) and weak nonlinearity (such that nonlinear events close to the boundary of the scattering volume can be neglected). Note that these two conditions are consistent with one another in the following sense: for larger optical thickness, a weaker nonlinearity is required to induce a comparable effect on the coherent backscattering cone. 

\section{Proof of the relation (\ref{rel2}) between cooperon and diffuson cross sections}
\label{proofrel2}

In this appendix, we will proof the equations
\begin{eqnarray}
{\rm Re} \left.\tilde{\kappa}\right|_{I_c=0} &  = & \frac{I_d}{2}\left.\frac{d}{dI_d}\sigma^{(d)}\right|_{I_c=0}\label{b1}\\
\left. {\rm Re}\{\tau^*K\}\right|_{I_c=0} & = & -\frac{I_d}{2}\sigma^{(d)}\frac{d}{dI_d}\left.\frac{1}{\ell}\right|_{I_c=0}\label{b2}
\end{eqnarray}
from which Eq.~(\ref{rel2}) results according to the definition of $\tilde{\sigma}$,
see Eq.~(\ref{tildesigma}).

To show Eq.~(\ref{b1}), we start with Eq. (\ref{tildekappa}) for $\tilde{\kappa}$, take its real part, and
express the expectation value $\langle\dots\rangle$ as an integral over the intensity distribution function $P(I)$, see Eq.~(\ref{pi}). Finally, we insert $f=Eg(I)$, and express the field derivatives $d/dE$ and $d/dE^*$ as intensity derivatives $d/dI$. Thereby, we obtain:
\begin{eqnarray}
{\rm Re}\tilde{\kappa} & = & {\mathcal N} \int_0^\infty dI P(I) \frac{d}{dI}I^2\frac{d}{dI}|g(I)|^2\\
& = & {\mathcal N}\int_0^\infty dI \left(P'(I)+\frac{I}{2}P''(I)\right) I|g(I)|^2
\end{eqnarray}
where we have used partial integration. Since Eq.~(\ref{b1}) is evaluated at $I_c=0$, we take
$P(I)=\exp(-I/I_d)/I_d$, see Eq.~(\ref{pi}) with $I_c=0$. 

In a similar way, we obtain
\begin{equation}
\sigma^{(d)}  =  -{\mathcal N}\int dI P'(I) I|g(I)|^2\label{b5}
\end{equation}
From this, Eq.~(\ref{b1}) follows with help of 
\begin{equation}
-\frac{I_d}{2}\frac{d}{dI_d}P'(I)=P'(I)+\frac{I}{2}P''(I)
\end{equation}
valid for $P(I)=\exp(-I/I_d)/I_d$.

To show Eq.~(\ref{b2}), we first note $K={\mathcal N}\int dI P(I) I|g|^2$, see Eq.~(\ref{k}), and
$P(I)=-I_dP'(I)$. Using Eq.~(\ref{b5}), we see that
Eq.~(\ref{b2}) is equivalent to
\begin{equation}
\left. {\rm Re}\tau^*\right|_{I_c=0}=\left. -\frac{1}{2}\frac{d}{dI_d}\frac{1}{\ell}\right|_{I_c=0}\label{b7}
\end{equation}
Now, we rewrite Eq.~(\ref{tau}) as
\begin{eqnarray}
\tau^*& = & \frac{i{\mathcal N}}{2k} \int_0^{\infty}dI P(I)\frac{d}{dI}I\frac{d^2}{(dI)^2}Ig(I)\\
 & = & -\frac{i{\mathcal N}}{2k}\int_0^{\infty}dI\left(IP'''(I)+2P''(I)\right)Ig(I)\label{b9}
\end{eqnarray}
On the other hand, Eqs.~(\ref{index},\ref{ell}) yield
\begin{eqnarray}
\frac{1}{\ell} & = & {\rm Im}\left\{\frac{\mathcal N}{k}\int_0^\infty dIP(I)\frac{d}{dI} Ig(I)\right\}\\
& = & -\frac{\mathcal N}{k}\int_0^\infty dI P'(I) I {\rm Im}g(I)\label{b11}
\end{eqnarray}
Finally, Eq.~(\ref{b7}) results from Eqs.~(\ref{b9},\ref{b11}) and
\begin{equation}
\frac{d}{dI_d}P'(I)=IP'''(I)+2P''(I)
\end{equation}
for $P(I)=\exp(-I/I_d)/I_d$.

We note that Eq.~(\ref{b1}) refers to nonlinear scattering, and Eq.~(\ref{b2}) to nonlinear propagation diagrams. Since the equality holds independently in both cases, the above derivation, explicitly performed for the case of nonlinear point scatterers, see Eq.~(\ref{model1}), can be generalized to other nonlinear models,
where scattering and propagation have different physical origin. In particular, this is the case in our second model, i.e. linear scatterers embedded in a homogeneous Kerr nonlinearity, see Eq.~(\ref{hom}).


\begin{thebibliography}{99}

\bibitem{anderson58}P. W. Anderson, Phys. Rev. {\bf 109}, 1492 (1958).
\bibitem{langer66}J. S. Langer and T. Neal, Phys. Rev. Lett.~{\bf 16}, 984 (1966).
\bibitem{kuga}Y. Kuga and A. Ishimaru, J. Opt. Soc. Am. A {\bf 1}, 831 (1984).
\bibitem{albada}M. P. van Albada and A. Lagendijk, Phys. Rev. Lett. {\bf 55}, 2692 (1985).
\bibitem{wolf}P. E. Wolf and G. Maret, Phys. Rev. Lett. {\bf 55}, 2696 (1985).
\bibitem{bec}D. Cl\'ement \textit{et al.}, Phys. Rev. Lett {\bf 95} 170409 (2005); C. Fort \textit{et al.}, Phys. Rev. Lett. {\bf 95} 170410 (2005); T. Schulte \textit{et al.}, Phys. Rev. Lett. {\bf 95} 170411 (2005); L. Sanchez-Palencia  \textit{et al.}, Phys. Rev. Lett. \textbf{98}, 210401 (2007) 
\bibitem{skipetrov}S.E. Skipetrov and R. Maynard, Phys. Rev. Lett. {\bf 85}, 736 (2000).
\bibitem{spivak}B. Spivak and A. Zyuzin, Phys. Rev. Lett. {\bf 84}, 1970 (2000).
\bibitem{altshuler99} I.L.~Aleiner,  B.L.~Altshuler, and M.E.~Gershenson, Waves Random Media \textbf{9}, 201 (1999).
\bibitem{wellens08}T.~Wellens and B.~Gr\'emaud, Phys.~Rev.~Lett. \textbf{100}, 033902 (2008).
\bibitem{wellens09}T. Wellens, Appl. Phys. B {\bf 95}, 189 (2009).
\bibitem{wellens05} T. Wellens, B. Gr{\'e}maud, D. Delande, and C.
  Miniatura, Phys. Rev. E {\bf 71} 055603(R) (2005); Phys. Rev. A {\bf 73} 013802 (2006)
\bibitem{wellens06b} T. Wellens and B. Gr{\'e}maud,
J. Phys. B: At. Mol. Opt. Phys. {\bf 39}, 4719 (2006)
\bibitem{agranovich91} V. M. Agranovich and V. E. Kravtsov, Phys. Rev. B \textbf{43}, 13691 (1991).
\bibitem{heiderich95} A. Heiderich, R. Maynard, and B. A. van Tiggelen, Opt. Comm. \textbf{115}, 392 (1995).
\bibitem{hartung08}M. Hartung, T. Wellens, C. A. M\"uller, K. Richter, and P. Schlagheck,
Phys. Rev. Lett. {\bf 101}, 020603 (2008).
\bibitem{paul05}T. Paul, P. Leb\oe uf, N. Pavloff, K. Richter, and P. Schlagheck, Phys. Rev. A \textbf{72}, 063621 (2005).
\bibitem{boyd}R. W. Boyd, {\it Nonlinear Optics} (Academic, San Diego, 1992).
\bibitem{gremaud08}B.~Gr\'emaud and T. Wellens, arXiv:0809.4533 (2008).
\bibitem{ishimaru}A. Ishimaru, {\it Wave Propagation and Scattering in Random Media} (Academic, New York, 1978), Vols. I and II.
\bibitem{wonderen}A. J. van Wonderen, Phys. Rev. B {\bf 50}, 2921 (1994).
\bibitem{goodman}J. W. Goodman, J. Opt. Soc. Am. {\bf 66}, 1145 (1976).
\bibitem{spivak03}B. Spivak and A. Zyuzin, J. Opt. Soc. Am. B {\bf 21}, 177 (2003).
\bibitem{chantry87}P. J. Chantry, J. Appl. Phys. {\bf 62}, 1141 (1987).




\end{thebibliography}
\end{document}